\documentclass[journal,twoside,web]{ieeecolor}
\usepackage{jsen}
\usepackage{cite}
\usepackage{amsmath,amssymb,amsfonts}
\usepackage{algorithmic}
\usepackage{graphicx}
\usepackage{textcomp}
\usepackage{wrapfig}
\usepackage{orcidlink}
\usepackage{tikz}
\usepackage{longtable}
\usepackage{array}

\usetikzlibrary{shapes.geometric, arrows}
\tikzstyle{startstop} = [rectangle, rounded corners, minimum width=3cm, minimum height=1cm,text centered, draw=black, fill=red!30]
\tikzstyle{process} = [rectangle, minimum width=3cm, minimum height=1cm, text centered, draw=black, fill=orange!30]
\tikzstyle{arrow} = [thick,->,>=stealth]

\def\BibTeX{{\rm B\kern-.05em{\sc i\kern-.025em b}\kern-.08em
    T\kern-.1667em\lower.7ex\hbox{E}\kern-.125emX}}
\definecolor{abstractbg}{RGB}{255,255,255}

\definecolor{ssred}{RGB}{230,193,194} 
\definecolor{ssblue}{RGB}{194,213,233} 

\begin{document}
\title{From Concept to Implementation: Streamlining Sensor and Actuator Selection for Collaborative Design and Engineering of Interactive Systems}
\author{%
  İhsan Ozan Yıldırım\orcidlink{0000-0002-2432-147X}, \IEEEmembership{Student Member, IEEE},
  Ege Keskin\orcidlink{0000-0002-5684-2229},
  Yağmur Kocaman\orcidlink{0000-0003-1595-8679},
  Murat Kuşcu\orcidlink{0000-0002-8463-6027}, \IEEEmembership{Member, IEEE}
  and Oğuzhan Özcan\orcidlink{0000-0002-4410-3955}
  \thanks{%
    İ. O. Yıldırım and E. Keskin are with Arçelik A.Ş. R\&D Sensor Technologies Directorate, İstanbul 34912, Türkiye (e-mail: \href{mailto:ihsanozan.yildirim@arcelik.com}{ihsanozan.yildirim@arcelik.com}, \href{mailto:ege.keskin@arcelik.com}{ege.keskin@arcelik.com}).%
  }
  \thanks{%
    Y. Kocaman is with Koç University - Arçelik Research Center for Creative Industries, Koç University, İstanbul 34450, Türkiye (e-mail: \href{mailto:ykocaman19@ku.edu.tr}{ykocaman19@ku.edu.tr}).%
  }
  \thanks{%
    M. Kuşcu is with the Electrical and Electronics Engineering Department, Koç University, İstanbul 34450, Türkiye (e-mail: \href{mailto:mkuscu@ku.edu.tr}{mkuscu@ku.edu.tr}).%
  }
  \thanks{%
    O. Özcan is also with Koç University - Arçelik Research Center for Creative Industries, Koç University, İstanbul 34450, Türkiye (e-mail: \href{mailto:oozcan@ku.edu.tr}{oozcan@ku.edu.tr}).%
  }
}

\IEEEtitleabstractindextext{%
\fcolorbox{abstractbg}{abstractbg}{%
\begin{minipage}{\textwidth}
\begin{wrapfigure}{r}{0.45\textwidth}%
\includegraphics[width=0.45\textwidth]{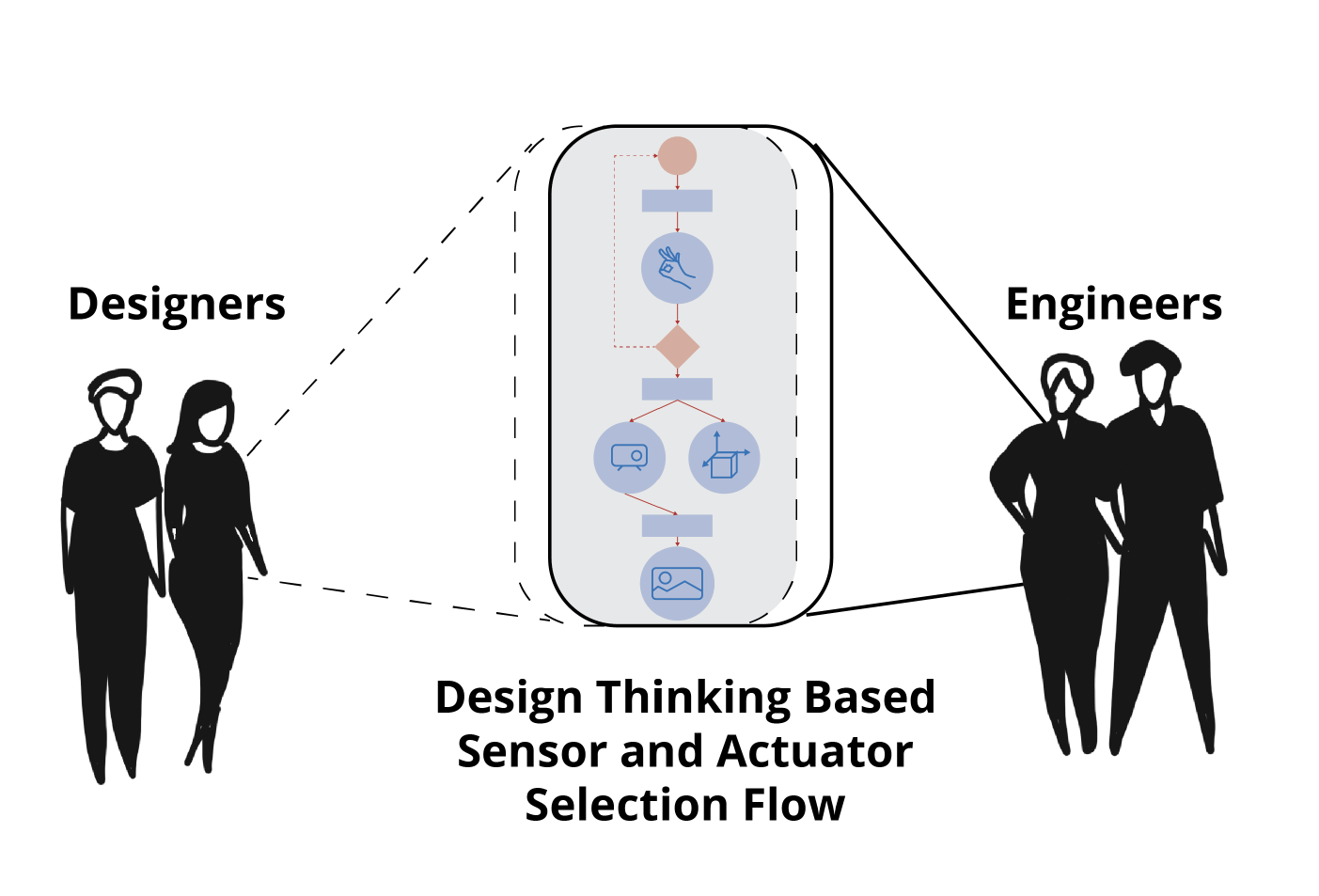}%
\centering
\end{wrapfigure}%
\begin{abstract}
Selecting appropriate sensors and actuators is a pivotal aspect of design and engineering, particularly in projects involving interactive systems. This article introduces the Design Thinking Based Iterative Sensor and Actuator Selection Flow, a structured decision-making approach aimed at streamlining this essential, yet often complex task. Created to accommodate individuals with diverse levels of technical expertise, our approach is uniquely suited for interdisciplinary teams of designers and engineers. Through the application of the flow to four real-world case studies, we highlight its broad applicability and demonstrate its efficacy in expediting project timelines and enhancing resource utilization. Our work lays a foundation for a more streamlined and user-centered process in selecting sensors and actuators, significantly benefiting the practice of interactive system design. This contribution serves as a seminal foundation for future research, offering significant contributions to both academic inquiry and practical applications across various industries. While the focus of the flow is on streamlining the selection process rather than on in-depth technical considerations, which are beyond the scope of this study, it provides a comprehensive guide for efficient and informed decision-making in the realm of interactive system design.

\end{abstract}

\end{minipage}}}

\maketitle

\section{Introduction}


Product developers, researchers, engineers and designers often find themselves in projects requiring them to implement prototypes that can vary in functionality and fidelity. Whether it is a working, semi-working or a Wizard of Oz (WoZ) \cite{kelley1984iterative} prototype, some degree of competence is usually expected when speaking in technical terms or proof of concept. A key challenge for prototyping is selecting appropriate hardware components, such as sensors or actuators, that can enable the system's desired input and output modalities. 

This article seeks to answer the research question: \emph{“How might teams of collaborating designers and engineers efficiently select sensors or actuators for their interactive system design and prototyping projects?”}. Answering this question is of paramount importance for the field of interactive system design. It provides a systematic and user-centered approach for teams of designers and engineers to streamline the selection process of sensors and actuators. This accelerates project timelines, optimizes resource utilization, and promotes efficient and informed decision-making, ultimately developing more effective and user-friendly interactive systems. We present the background motivation for this question and elaborate on its scope. The research question is inspired by the following observations.

\begin{figure}
    \centering
    \includegraphics[width=1\linewidth]{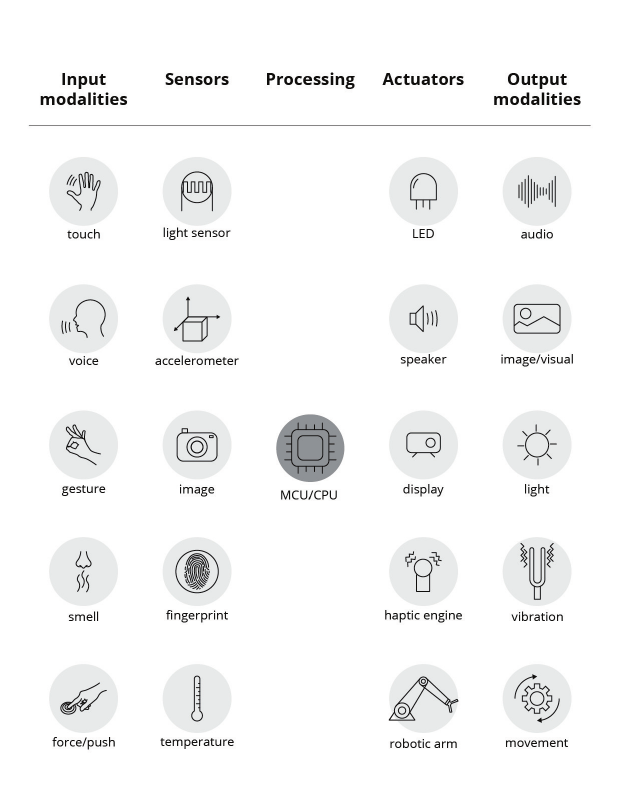}
    \caption{Interactive system components with examples, input and output modalities as physical phenomenons, sensors, actuators and processing devices.}
    \label{fig:intsys}
\end{figure}

The concept of interaction refers to the process by which entities influence or change each other's actions over time. In the context of our study, these entities are humans and digital systems \cite{hornbaek2017interaction}. An interactive system serves to aid the user in accomplishing goals from some area of expertise in a real-world activity. It consists of goals, tasks, intentions, the system’s and the user’s language, and the system itself, which is a computerized application \cite{HCI-dix}. When designing interactive systems, designers and engineers often endeavor how users come into contact (direct and intentional, or indirect and unintentional) with an artifact and receive a response in the form of feedback. The input from the user is typically a physical phenomenon, such as speech, touch, audio, or a magnetic field, that is digitized by various sensors, see Figure \ref{fig:intsys}. Once this input is digitized, it can be processed to generate an appropriate output through actuation. Examples of such outputs are sound, light, movement, or alike \cite{HCI-dix}. For instance, when designing a user interface, interaction designers decide on which input and output modalities would best fit the users' context in order to meet the predefined user goals. Since most interaction problems, especially human-computer interaction problems rely on user input and feedback mechanisms, sensors and actuators are often selected together to enable both sides of the interaction loop. 

The selection of appropriate sensors and actuators becomes particularly challenging for non-technical designers \cite{Varshney2021, Algamili2021}, often investing a significant amount of time and effort in this phase. This not only slows down the design process but can also lead to a wide range of sub-optimal choices or an overwhelming array of alternatives, ultimately compromising the project goals. 

In light of the challenges described, this article proposes a Design Thinking based flow that facilitates the selection process of sensors and actuators for interactive systems. The flow, which draws inspiration from the iterative and user-centred approaches of the Design Thinking methodology \cite{Stackowiak2020}, is designed to be mainly applied in the ideation and prototyping phases of Design Thinking processes. It consists of several blocks that follow each other, creating a step wise foundation for streamlining the practice of comprehensive prototyping. The flow is intended to be followed sequentially without skipping any step; however, it also allows for iterations to previous if the outputs of those steps are found to be incomplete or inaccurate, or if there is room for improvement. This contribution is expected to benefit a wide audience, from individuals with limited technical expertise in sensors and actuators to experienced designers and engineers working in collaborative and interdisciplinary teams.

The proposed flow was iteratively developed and fine-tuned by building on our experiences across various projects at the intersection of interaction design, engineering, and design research. To demonstrate its applicability and benefits, we present four of these practical projects, where this approach was employed, each differing in scope, context, and objectives. Whether it is designing consumer goods with interactive features or conducting a proof-of-concept study on gesture recognition, our approach has proven to be both flexible and broadly applicable. It is important to clarify that the aim of this article is not to discourage readers from diving into the technical details of sensors and actuators. Instead, we focus on streamlining the selection process. While we do not discuss technical aspects such as digital communication protocols or specific hardware capabilities, the guidelines here provide a foundational understanding of how to approach the selection process itself. As such, this study can also support collaboration and communication between professionals with different backgrounds and expertise, such as designers seeking engineering advice or engineers in need of design perspectives.

\section{Background and Related Work}

Within our study, we navigate through existing and well-established approaches in previous research. This exploration is crucial as our proposed flow's foundation is constructed upon the inspiration and insights garnered from these methodologies. Notable methodologies in the field include Design Thinking \cite{Stackowiak2020,brown2008design}, which has played a significant role in shaping collaborative design and engineering. Our specific focus revolves around designers collaborating with engineers, with the goal of presenting a practical, tailored flow for technology selection that caters to their unique needs.

Existing research and studies in the field primarily concentrate on the technical aspects and performance of sensors and actuators, often presupposing a certain level of expertise from the user \cite{li_aoi-based_2023, zhu_optimal_2019, maddikunta_industry_2022}. Notably, our observation reveals that the majority of publications in this domain date back at least a decade. This temporal gap suggests a relative lack of recent attention to the topic, signaling the need for further exploration and clarification in the context of sensor and actuator technology selection.

In our pursuit of relevant literature, we employed two distinct approaches: keyword search and backward search. The keyword search method involved the input of specific terms related to sensor and actuator technology selection into search engines or databases, enabling the evaluation of the relevance of each result \cite{liu2006effective}. On the other hand, the backward search, also referred to as backward citation chasing, entailed examining articles that have cited specific studies in the domain \cite{haddaway2022citationchaser}. Although a valuable method, backward search poses challenges as not all databases provide a comprehensive history of the works citing a particular study. Additionally, citations cannot be continually updated in an offline document, distinguishing them from references that are documented once. This dual-method approach ensured a thorough and multifaceted exploration of the existing literature on sensor and actuator technology selection.

Design Thinking, rooted in a rich history and iterative problem-solving methodologies, stands as a prominent approach in our exploration. Originating in the field of design, it has evolved into a versatile framework widely applied in various domains, including interactive systems and technology selection. The significance of Design Thinking lies in its human-centered and collaborative nature, fostering empathy for end-users and encouraging multidisciplinary collaboration \cite{brown2010design}. Initially gaining traction in design and architecture, Design Thinking's principles have transcended disciplinary boundaries, becoming instrumental in solving complex problems and fostering innovation \cite{lockwood2009design}. Its emphasis on understanding user needs, ideation, prototyping, and continuous refinement aligns with the iterative nature of technology design. Well-established methodologies such as the Double Diamond model \cite{kochanowska2021double} and Stanford's d.school framework provide structured paths for Design Thinking application \cite{stanfordDschool}. Design Thinking is not limited to a single prescribed approach; rather, it accommodates diverse methodologies, ensuring adaptability to various contexts \cite{gaulton_how_2023, bartoloni_towards_2022}. This versatility positions Design Thinking as a valuable foundation upon which our proposed flow for sensor and actuator technology selection is constructed, allowing for a human-centric and collaborative approach tailored to the specific needs of designers collaborating with engineers.

\begin{figure}
    \centering
    \includegraphics[width=1.1\linewidth]{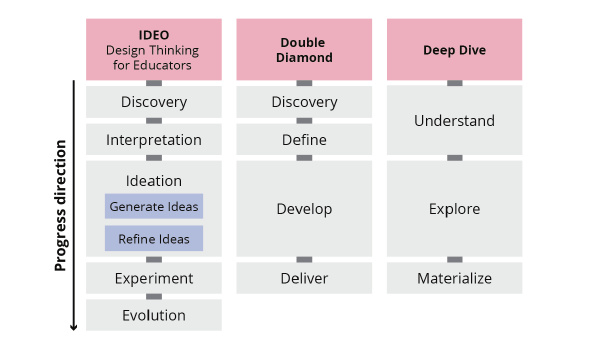}
    \caption{A schematic representation of different Design Thinking methodologies in the literature from left to right respectively \cite{ideoDesignThinkingEducators}\cite{kochanowska2021double}\cite{stanfordDschool}.}
    \label{fig:lit_flow}
\end{figure}

In the realm of Design Thinking, various methodologies have emerged, each offering unique perspectives and structured processes to guide problem-solving and innovation. The Double Diamond model, proposed by the Design Council, embodies the iterative nature of the Design Thinking process, comprising four key stages: Discover, Define, Develop, and Deliver \cite{kochanowska2021double}. This model advocates for a divergent-convergent approach, encouraging broad exploration of problem spaces followed by focused problem definition and solution development. Stanford's d.school, a pioneer in applying Design Thinking to education, provides a structured framework with five key stages: Empathize, Define, Ideate, Prototype, and Test \cite{stanfordDschool}. This approach places a strong emphasis on understanding user needs, ideation, and prototyping as integral components of the design process. IDEO, a renowned design and innovation consultancy, has introduced methodologies such as the Deep Dive and IDEO's Inspire, Ideate, Implement. The Deep Dive model emphasizes phases such as Understand, Explore, and Materialize \cite{brown2008design}. IDEO's Inspire, Ideate, Implement approach, as well as their Design Thinking for Educators model (Discovery, Interpretation, Ideation, Experimentation, Evolution), further contribute to the diverse landscape of Design Thinking methodologies \cite{ideoInspireIdeateImplement, ideoDesignThinkingEducators}. Figure \ref{fig:lit_flow} shows the alignment of the schematic representations of three popular examples. Such methodologies have been successfully applied in diverse contexts, from healthcare \cite{gaulton_how_2023} to technology innovation \cite{bartoloni_towards_2022}, showcasing the adaptability and effectiveness of Design Thinking principles. As we delve into the sensor and actuator technology selection process, drawing inspiration from these established methodologies and their alignment, we aim to integrate human-centric and collaborative approaches, ensuring a comprehensive and inclusive decision-making process tailored to the needs of both designers and engineers.

Beyond the established Design Thinking methodologies, recent research has explored innovative workflows that merge Design Thinking with other approaches or extend its application into diverse contexts. For instance, Gaulton et al. present a "Human-Centered Quality Improvement" approach in the perinatology domain, integrating Design Thinking and Quality Improvement to address complex problems \cite{gaulton_how_2023}. Bartoloni et al. propose a conceptual model, applying Design Thinking to the Quintuple Helix innovation framework, fostering effective integration of technology into society, as demonstrated in a healthcare project called SMARTAGE \cite{bartoloni_towards_2022}. Ahmad et al. explore the convergence of 6G communication, extended reality, and IoT big data analytics for healthcare, providing a comprehensive review that emphasizes the interplay of these technologies in shaping future healthcare systems \cite{ahmad_leveraging_2023}. Tsai et al. contribute an empirical study demonstrating how Design Thinking with constructivist learning increases learning motivation and wicked problem-solving capability in the context of education in Taiwan \cite{tsai_design_2023}. These studies showcase the versatility of Design Thinking, not only as a standalone methodology but also as an integrative force that collaborates with other frameworks to address challenges in various domains. As we navigate the landscape of sensor and actuator selection, we draw inspiration from this rich tapestry of methodologies to craft a unique flow that harmonizes the strengths of Design Thinking with the specific demands of our collaborative design and engineering context.

Sensor and actuator selection might mean different things in different contexts, often entailing various dimensions of decision-making. Some contexts may focus on selecting which sensor or sensors' data to use in a sensor network, while others revolve around choosing a specific sensor or actuator model based on parameters like cost, energy consumption, or size. However, our specific focus diverges from these conventional interpretations. Instead, our emphasis is on the strategic and collaborative decision-making process involved in selecting the most fitting sensor and actuator technologies. This process is not merely about technical specifications or operational parameters; rather, it extends into the realm of interactive system design, where the chosen technologies align with the unique needs of the system and user requirements. Our approach takes inspiration from existing methodologies and design thinking principles, emphasizing the importance of critical design considerations and user-centric perspectives in the sensor and actuator selection process \cite{liu2006effective,7487646,s21196470,4663892,10.1117/12.723514,10.1145/1127777.1127783,shamaiah2010greedy,zappi2008activity,li_aoi-based_2023}.

Previous research has predominantly focused on the technical intricacies of sensor and actuator technologies, often assuming a high level of expertise from the user. Notably, we found that recent studies have explored the use of machine learning techniques to assist in sensor selection \cite{a1020130}. In the realm of robotics, the importance of proper sensor selection was underscored through comparative analyses, demonstrating the impact of appropriate sensor choices on system performance \cite{7487646}. Mathematical derivations were employed to justify the suitability of specific sensor types.

Another facet of sensor and actuator selection is the consideration of factors such as cost, performance, and operating conditions. Some works have utilized charts to facilitate balanced decision-making \cite{shieh2001selection}, while others have proposed frameworks for optimal sensor selection in specific environments, accounting for factors like failure rates and system downtime \cite{s21196470, 4663892}.

As we delve into sensor network applications, particularly in settings like smart homes, technical surveys have been conducted to address aspects like coverage and localization techniques \cite{10.1117/12.723514}. Energy consumption becomes a crucial consideration, with studies emphasizing the prioritization of sensor selection based on the benefit provided versus the energy consumed \cite{10.1145/1127777.1127783}.

In our case, we draw insights from sensor and actuator selection methodologies to develop a streamlined approach that facilitates collaboration between non-technical designers and their technical counterparts, ensuring effective technology selection for interactive system design. Our review highlights a prevailing emphasis on technical and engineering considerations in existing literature, often overlooking critical design aspects and user needs. This observation underlines the motivation for our research inquiry: How can we develop a streamlined approach that empowers non-technical designers to collaborate effectively with their technical counterparts in selecting sensor and actuator technologies?

\section{Design Thinking Based Iterative Sensor and Actuator Selection Flow} 

This article introduces a Design Thinking-based, iterative flow designed for the efficient selection of sensor and actuator technologies in the development of interactive systems. Tailored for project teams comprising both designers and engineers collaborating on projects involving the creation of interactive systems capable of sensing and manipulating the physical world, the flow encompasses seven distinct steps. Each step corresponds to a crucial phase in the design process, aligning with established Design Thinking methodologies.

The initial two steps of the proposed flow align with the "Inspire" phase of IDEO's methodology \cite{ideoInspireIdeateImplement}. In these steps, the team is tasked with extracting or discovering the affordances inherent in the design problem. Subsequently, these affordances are interpreted to define the interactions within the system. The remaining four steps of the flow correspond to the "Ideate" phase of IDEO's approach. Here, the team generates and refines ideas related to sensor and actuator selection. The culmination of this phase involves making the final decision on the sensor and actuator technologies. Additionally, teams may choose to prototype the interactive system and iterate through the flow based on preferences or emerging insights. This step aligns with the "Implement" phase of IDEO's methodology.

\begin{figure}
    \centering
    \includegraphics[width=1\linewidth]{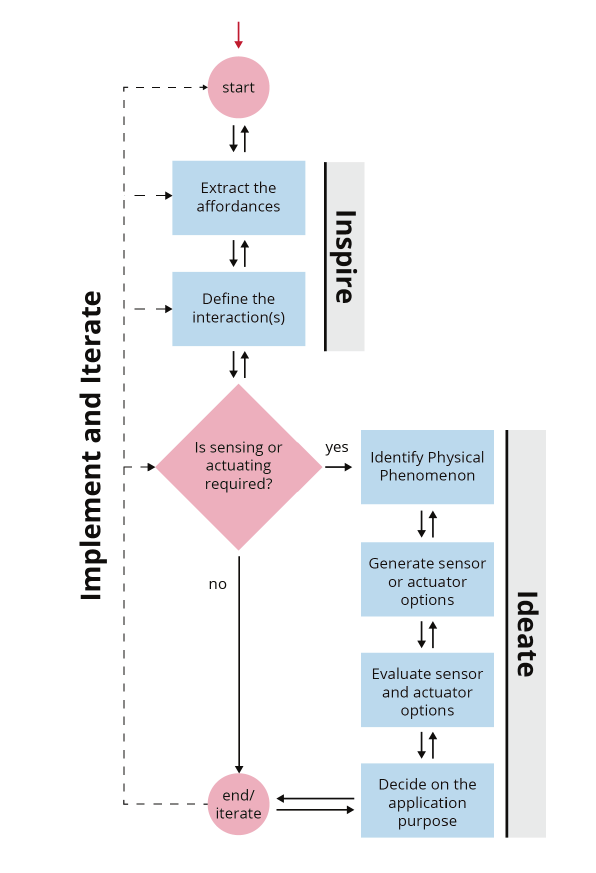}
    \caption{Overview of the Design Thinking-based iterative sensor and actuator selection flow, demonstrating its alignment with IDEO's Ispire, Ideate, Implement methodology.}
    \label{fig:flowchart}
\end{figure}

The proposed flow is inherently iterative and adaptable, allowing for flexibility in addressing diverse design challenges. In the subsequent sections, we provide a comprehensive explanation of each step in the flow and offer practical insights into its application through the illustration of four example projects we have conducted. Figure \ref{fig:flowchart} visually depicts an overview of the proposed Design Thinking-based iterative sensor and actuator selection flow, showcasing its alignment with a popular Design Thinking methodology from IDEO \cite{ideoInspireIdeateImplement}.

The flow consists of the following steps:

\textbf{1. Extract the Affordance(s):} In the first step, the team clarifies the design problem by identifying what affordances they want to provide to their users. Affordances are ``the perceived and actual properties of a thing that determine how it can be used" \cite{norman2013design}. By extracting affordances from vague or complex problems, such as `wicked problems' \cite{buchanan1992}, the team can focus on what they want their system to do rather than how they want it to do it. 

\textbf{2. Define the Interaction(s):} The term interaction can be defined as entities manipulating or altering each other's behaviour as time passes \cite{hornbaek2017interaction}. In our case, entities are digital systems and humans. Building upon the clarified design problem and identified affordances, the team refines the well-defined problem identified in the previous step by specifying the distinct interactions that the interactive system should facilitate. This entails elucidating how users will engage with the system and outlining the corresponding actions the system is expected to undertake. To inform this step, relevant interaction types \cite{HCI-dix} provide valuable insights into establishing the nature of user-system engagements. By defining these interactions, the team gains a comprehensive understanding of the desired functionality and behavior that the system should embody. For instance, in the context of a smart jewelry piece, this step might involve detailing how the jewelry should respond to various gestures, inputs, or conditions, providing a clear roadmap for subsequent sensor and actuator technology selection.


\textbf{3. Determine the Need for Sensing or Actuating:} Following the definition of interactions, the team critically assesses whether the desired interactions necessitate the incorporation of sensor or actuator technologies. If the interactions do not involve sensing or actuation, the team may explore alternative design approaches or determine that the problem falls outside the scope of this flow and article. 

\textbf{4. Identify Physical Phenomenon:} Once the team has determined that sensing or actuating is necessary, they move on to identifying the specific physical phenomena that need to be measured or manipulated to achieve the desired interactions. These physical phenomena could be anything from temperature, light intensity, motion, pressure, or any other measurable attribute. The identification of these phenomena serves as a fundamental step in selecting suitable sensor and actuator options. For example, if the system's objective is to monitor user movement, pinpointing the physical phenomenon of "motion" becomes pivotal in determining the appropriate sensor technology.

\textbf{5. Generate Sensor and Actuator Options:} With a clear understanding of the physical phenomena involved, the team generates a wide range of sensor and actuator options. This brainstorming process aims to explore various possibilities without constraints. Ideas can be inspired by existing technologies, prior experiences, or innovative concepts. These options might involve different sensor types, such as accelerometers, light sensors, temperature sensors, and actuator types, like motors or LEDs. The goal is to create a diverse pool of options that will be evaluated in the subsequent steps. This step encourages creativity and opens the door to novel solutions that align with the system's requirements.

\textbf{6. Evaluate Sensor and Actuator Options:} In this step, the team critically assesses each sensor and actuator option generated in the previous step. They evaluate the feasibility and suitability of each option based on predefined criteria. These criteria are specific to the project's requirements and objectives, ensuring that the chosen technologies align with the intended application. For instance, if the system needs high accuracy and low power consumption, the team might assign higher importance to these criteria during the evaluation process. Depending on the complexity and importance of the criteria, the team might employ quantitative metrics, qualitative ratings, or a combination of both to compare and rank the options. By evaluating each option thoroughly, the team can identify the best-fit technologies that will contribute to the success of the interactive system.

\begin{figure}
    \centering
    \includegraphics[width=1\linewidth]{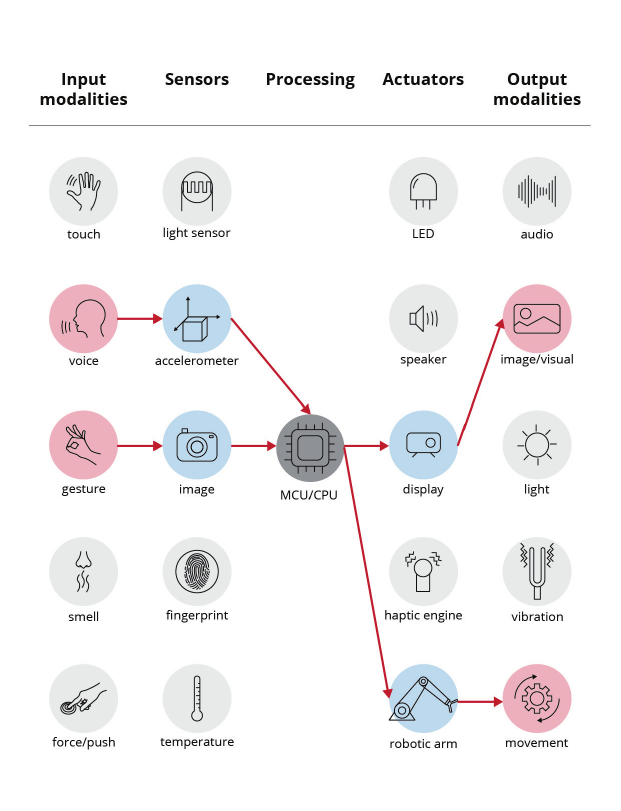}
    \caption{Identifying physical phenomenon (such as touch, speech, gestures, light, etc.) as input and/or output modalities (red circles) and selection (blue circles) of sensors (light sensors, accelerometers etc.) and actuators (haptics engines, LEDs, robotic arms etc.). The arrows depict the possible relationships between the components of the interactive system.}
    \label{fig:phenomenon-selection}
\end{figure}

\textbf{7. Decide on the Application Purpose:} With the sensor and actuator options thoroughly evaluated, the team proceeds to make a crucial decision regarding the purpose of the interactive system. Careful consideration is given to whether the system will be developed as a rapid prototype for swift user interaction and feedback, designed for mass production with adherence to specific standards, durability requirements, and production constraints, or positioned somewhere in between. This intermediate stage might involve iterative evaluations to refine the design before transitioning to a mass-production version or signify a research-oriented project with no immediate plans for mass-production. The decision significantly influences the overall development process, timeline, and resource allocation. By aligning the chosen purpose with the project's goals and constraints, the team ensures a strategic and successful realization of the interactive system.

\begin{figure}
    \centering
    \includegraphics[width=1\linewidth]{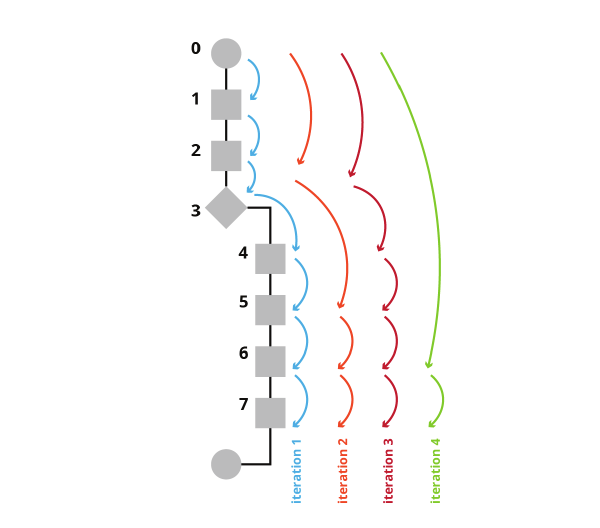}
    \caption{The flow's iterative behaviour allows for starting at any point and skipping steps.}
    \label{fig:iterate-general}
\end{figure}

Termed as "End/Iterate," the final step of this flow involves deciding whether to conclude the process or iterate to a previous step, including a potential return to the beginning. The dashed line in Figure \ref{fig:flowchart} aligns with the Implement, Experiment, or Evolve steps in recognized Design Thinking methodologies. At this juncture, the team may opt to implement or prototype the generated ideas, subject them to testing, gather user feedback, and proceed with refining the system. Alternatively, the team retains the flexibility to iterate to any preceding point in the flow as depicted in Figure \ref{fig:iterate-general}, even revisiting the initial stages, without necessarily undergoing prototyping. The sole constraint is the sequential nature of the steps, ensuring that no step is skipped, as each builds upon the preceding ones. This adaptability enhances the flow's effectiveness in accommodating diverse project needs and objectives.

We will illustrate the application of the flow through four real-world examples next and evaluate the results in the discussion section. However, it is worth going through a brief practical example of how one can link the bridge between input and output modalities. In Figure \ref{fig:phenomenon-selection}, let us assume a system is to be controlled with voice commands and hand movements. These may correspond to voice and gestural phenomena in the input modalities column. As the flow in Figure \ref{fig:flowchart} is followed, one might discover that it is possible to interpret gestures with image-sensing technologies and voice with an accelerometer (because sound creates vibration.). These measured quantities translate to numerical data. This data can then be further processed with a processing unit of some sort. Afterwards, the processing unit can trigger a display to show images on a screen, indicating the status of the system or manipulate a robotic arm to enact hand movements or speech. The resulting selection of one iteration is visually depicted in Figure \ref{fig:phenomenon-selection}.

\subsection{Autonomous Yoghurt Machine Project}

\subsubsection{Project Overview} The aim of this project is to create a fully automated yoghurt machine that can detect the acidity level of yoghurt. This goal was established based on user feedback, where they expressed a need for a device that could handle the entire yoghurt-making process, ensure a consistent taste, and prevent the yoghurt from becoming sour due to an incomplete process. The project team has formulated the following problem statement: \emph{"Producing consistent and desired quality yoghurt using different types of milk and cultures can be a challenging process."}

To solve the problem mentioned earlier, a team of engineers and industrial designers created a solution: an all-in-one, self-sufficient yoghurt maker designed to be a compact household appliance. The machine can be easily placed on a counter top or stored in cabinets. It takes partial control of the fermentation process, while the user is responsible for supplying the milk and culture mixture, and indicating the desired yoghurt taste level. The user just needs to mix the milk and culture beforehand, and the machine takes care of the rest. To get started, the user simply presses a button and waits for the process to end.

The team systematically explored various options for detecting yoghurt temperature, texture, and acidity within the iterative framework of the proposed flow. These options encompass diverse sensing methods such as pH measurement and optical transmissivity. However, given the project's overarching objective of mass production, the team diligently weighed the trade-offs between cost and functionality in their selection process. Certain sensing methods were excluded from consideration due to their prohibitively high cost, prompting the team to seek out alternative approaches that offered comparable effectiveness but were more cost-effective. The iterative approach allowed the team to navigate the decision-making process methodically, ensuring that the chosen sensing methods aligned with the project's technical requirements and budgetary constraints. 

\subsubsection{Application of the Design Thinking Based Iterative Flow}

\begin{figure}
    \centering
    \includegraphics[width=1\linewidth]{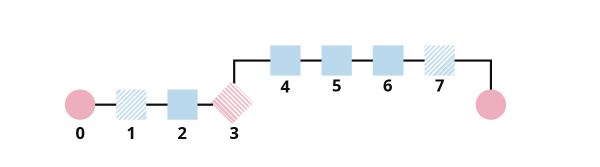}
    \caption{The schematic representation of the Iteration 4 of the flow during the development. Full shapes mean revisited or changed, and hatched boxes indicate being skipped or not changed steps.}
    \label{fig:yoghurt_it4}
\end{figure}

It is possible to observe the application of the proposed iterative flow in Table \ref{tab:ssi_yog12} and \ref{tab:ssi_yog34}. The text that appears in \textit{italics} indicates that there have been no changes made from the previous iteration. It is important to note that revisiting and making changes on every step during each iteration is not mandatory. However, it is crucial to be meticulous because the flow is progressive, and each step is built upon the previous one. The iterative behaviour of the flow could also be seen in the schematic representation of the fourth iteration in Figure \ref{fig:yoghurt_it4}.

\begin{table}
    \centering
    \caption{Application iterations (1,2) of the flow for autonomous yoghurt machine project}
    \includegraphics[width=1\linewidth]{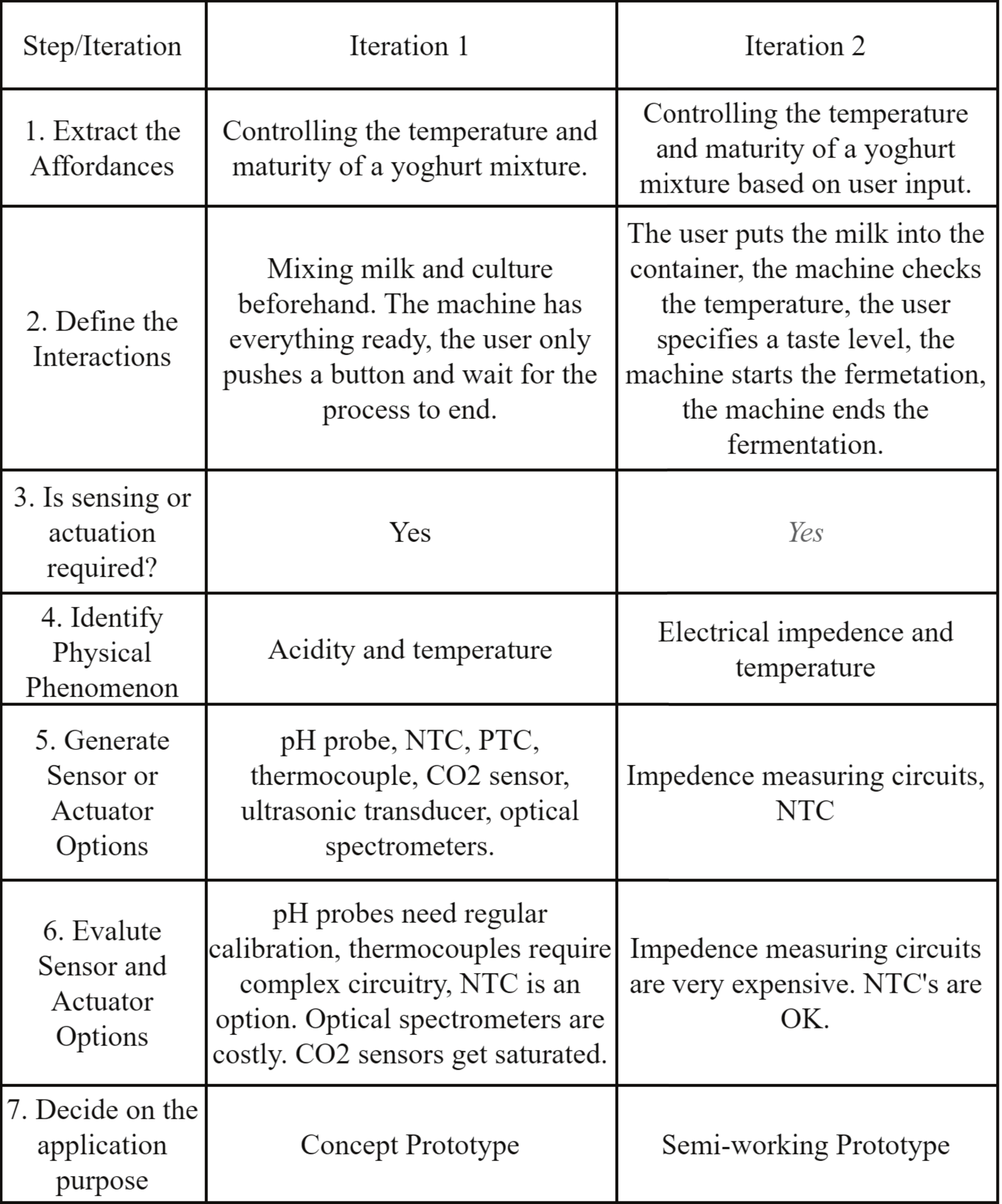}
    \textit{Italic} text represents no change compared to the previous iteration.
    \label{tab:ssi_yog12}
\end{table}

\begin{table}
    \centering
    \caption{Application iterations (3,4) of the flow for autonomous yoghurt machine project}
    \includegraphics[width=1\linewidth]{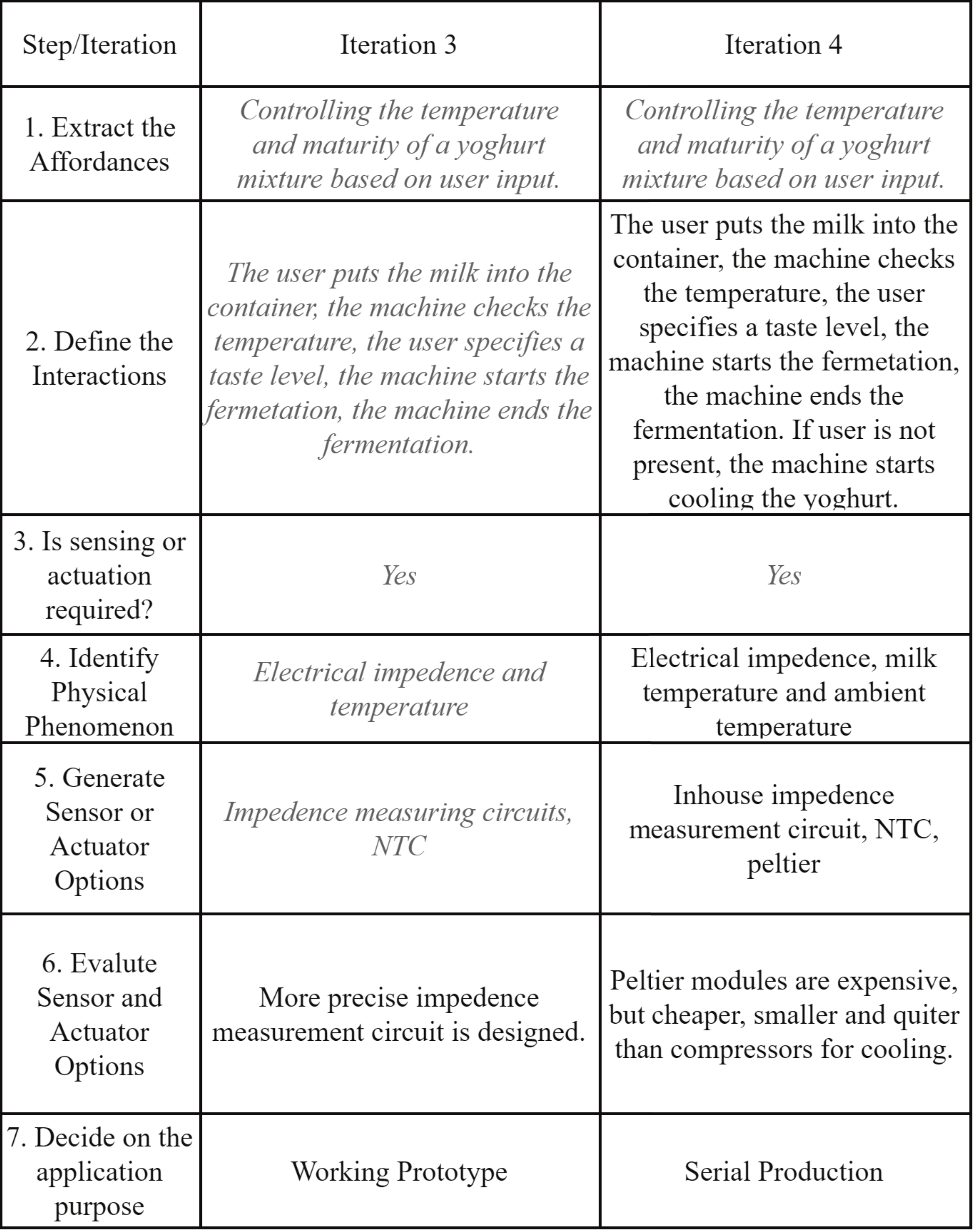}
    \textit{Italic} text represents no change compared to the previous iteration.
    \label{tab:ssi_yog34}
\end{table}

\paragraph{\textbf{Extract the affordance(s)}} 
The project was initiated due to a lack of knowledge on the user side regarding the process of making yoghurt at home. People often express uncertainty about the correct temperature for fermenting yoghurt and adding culture to milk. This uncertainty frequently leads to unsuccessful fermentation attempts, resulting in spoiled milk. Additionally, since every combination of milk and culture yields yoghurt with a different taste within a fixed amount of time, keeping track of the \emph{sourness} is challenging for individuals without technical training. A suitable problem statement for this case could be phrased as: \emph{"People experience difficulties while determining the correct temperature for fermenting yoghurt, and different milk/culture combinations make the fermentation process unpredictable."} The end goal should be to address the issues mentioned in the problem statement. Given the project's goal of mass production, it is imperative to specify the affordance rather than defining the exact product with its specifications. Otherwise, the solution will be biased towards predefined listings before the project even starts.

\paragraph{\textbf{Define the interaction(s)}}
We need to lay out possible interactions to address this problem strategically. This will allow us to consider different perspectives regarding user expectations. First and foremost, we need a way to check the temperature of the milk and monitor the change in the sourness of the milk/culture mixture during fermentation. Two interaction scenarios can be outlined for this case, although it is possible to develop more, and interactions have been refined through iterations as in Tables \ref{tab:ssi_yog12} and \ref{tab:ssi_yog34}.
\begin{itemize}
    \item The user prepares the milk and culture separately, puts the milk into the container, and the machine checks the temperature. It informs the user to add the culture if the temperature is suitable. After the user specifies a taste level, the machine initiates the fermentation process and concludes it when the sourness reaches the desired level. The user can then retrieve the yoghurt from the machine.
    \item The milk and culture are already stored in the machine. When the user decides to make yoghurt, all they have to do is turn it on and provide the desired taste level. The machine measures the sourness of the yoghurt and notifies the user, similar to the first interaction case.
\end{itemize}
The two cases primarily differ in the preparation phase. The second case offers a more all-in-one experience, while the first one involves the user more in the process, providing a more immersive experience. As a result of user studies dedicated to this issue, the first case was found to be more suitable, immersive, and feasible.

\paragraph{\textbf{Determine the need for sensing or actuating}}
Referring back to our problem statement, the user clearly faces challenges regarding the temperature of the milk and the acidity of yoghurt during fermentation. Both of these quantities need to be measured (or sensed) and validated by the machine we are designing. Similarly, the user needs to be notified when the fermentation is aborted by the machine. This aspect also involves a form of actuation. 

\paragraph{\textbf{Identify physical phenomenon}}
In this step, we consulted food engineers to comprehend which properties of yoghurt change throughout the fermentation process. Phrases such as: \emph{"How can I know when my yoghurt is ready?"} or \emph{"What are the most important steps to follow while making yoghurt?"} led us to identify acidity \cite{galster1991ph} and temperature \cite{oracc2019different} measurement as the most crucial physical/chemical properties of yoghurt during fermentation. Acidity is highly correlated with the pH of a substance \cite{sadler2010ph}. Consequently, our keywords for the next step in searching for sensors are \textbf{temperature sensor} and \textbf{pH sensor}. As iterations took place, physical phenomenon to be identified were narrowed down to electrical properties of yoghurt.

\paragraph{\textbf{Generate sensor and actuator options}}
Now that we have identified what needs to be measured, it's time to search for suitable sensors and actuators. We consulted electrical and mechatronics engineers with the phrases such as: \emph{"How can I measure temperature?"} and \emph{"How can I measure pH?"}, which yielded following sensor options: pH electrodes \cite{wang2002long}, pH sensors \cite{safavi2003novel}, NTC (Negative Temperature Coefficient) \cite{fagan1993reliability}, PTC (Positive Temperature Coefficient) \cite{nakamura2016development}, thermocouples \cite{ballantyne1977fine}, thermometers \cite{pearce2002brief}, digital thermometers \cite{hadley1991inexpensive}, and thermometers in IC (Integrated Circuit) form \cite{wu2019remote}. Similarly, some actuator options could be speakers (Audio actuator), buzzers (Audio actuator), LEDs, and digital screens (Visual actuators). The next step is to evaluate each of these options and proceed with the process of elimination.

\paragraph{\textbf{Evaluate sensor and actuator options}}

In this phase of the proposed flow, each previously listed sensor type was evaluated individually. The evaluation method and criteria are not mandated by the flow as they tend to be different by the nature of every project. Their strengths and weaknesses were assessed specifically for the yoghurt fermentation case, and their relevance to the subject was identified. Research among all of the options yielded the following results, these findings helped the team to evaluate sensor and actuator options during all four iterations:

\begin{itemize}
\item \textbf{pH Electrode:} A pH electrode is the first piece of equipment that comes up when researching pH measurement. Its working principle relies on electrochemical phenomenon, where the power of hydrogen in a liquid can be converted into a measurable voltage. Although they can provide a signal that is directly proportional to the pH value of a liquid, they need additional circuitry and hardware. This, combined with the need for constant calibration, made this option seem feasible at first, but not usable in a small domestic appliance later on. Iteration towards previous steps helped the project team to better focus their sensor options towards more feasible alternatives.

    \item \textbf{pH Sensor:} pH sensors are plug-and-play devices usually designed for laboratory use. They contain all the required equipment to provide a fast pH reading of a liquid. Due to their price point and form factor, they were not suitable for use in a yoghurt machine. 
     
    \item \textbf{NTC:} These devices change their resistance inversely proportional to the temperature. For this reason, they require minimal extra components. Due to their compact size, they can be encapsulated into various casings (needle type, ring mount, mushroom type, food grade) for different use cases. They were found suitable for the case of yoghurt fermentation.

    \item \textbf{PTC:} These devices change their resistance directly proportional to the temperature. For this reason, they require minimal extra components. They are mainly used as heaters \cite{musat2010characteristics} in certain appliances. Although temperature measurement is possible with PTCs, their casings are not as flexible as the NTCs. This is the reason PTCs were declined as a temperature measurement solution. 

    \item \textbf{Thermocouple:} Thermocouples consist of two metals with different materials that are joined together in a pair \cite{adhikari2017thermocouple}. A voltage drop that is proportional to the joint temperature is produced across the pair, which can then be amplified and used to calculate temperature. Unlike NTCs, thermocouples require more demanding hardware to work. Combined with relying heavily on the joint quality, they were dismissed as a temperature measurement tool. 

    \item \textbf{Thermometer:} Analog thermometers do not provide any signal that can be processed electrically. For this reason, they were dismissed as a temperature measurement tool.

    \item \textbf{IC Thermometer:} There are digital thermometers in the form of tiny ICs that can be mounted on circuit boards. They are usually used to monitor air quality and are not submersible in liquids. They were dismissed as a temperature measurement tool. 

    \item \textbf{Digital Thermometer:} Digital thermometers are similar to regular analogue thermometers. They usually provide an intuitive display for easier visualization of the temperature. Since they cannot provide the temperature information in any other way than through a screen, they were declined as a solution for temperature measurement during yoghurt fermentation.
\end{itemize}

The above findings were summarized in Table \ref{table:1}

\begin{table}
\centering
\caption{Summary of overall sensors and evaluations}
\small
\begin{tabular}{|c c c c|}
\hline
\textbf{Sensor} & \rotatebox{90}{\textbf{Measured Quantity}} & \rotatebox{90}{\textbf{Initial Decision}} & \rotatebox{90}{\textbf{Final Decision}}\\
\hline
pH Electrode & pH & Go & Not go\\
pH Sensor & pH & Not go & Not go\\
NTC & Temp. & Go & Go\\
PTC & Temp. & Go & Not go\\
Thermocouple & Temp. & Not go & Not go\\
Thermometer & Temp. & Not go & Not go\\
IC Thermometer & Temp. & Go & Not go\\
Digital Thermometer & Temp. & Not go & Not go\\
\hline
\end{tabular}

\label{table:1}
\end{table}

As you have hopefully noticed, both of the pH measurement methods have been eliminated. At this step, an iteration was necessary to look for alternative methods to sense the pH of yoghurt indirectly. A literature search with the phrase \emph{"Indirect pH measurement of yoghurt"} yielded a variety of options and methods that establish an indirect correlation between the state of the fermentation and some physical or chemical phenomenon. Some options were:

\begin{itemize}
\item Electrical impedance measurement \cite{bodor2020monitoring}
\item Optical spectroscopy \cite{yakubu2022recent}
\item Optical scattering \cite{arango2020inline}
\end{itemize}

By following the proposed flow, the options were significantly narrowed down to three items to be more specific. The related research and development phases began once this flow was completed. 

\paragraph{\textbf{Decide on the application purpose}}
The project aimed for mass production, necessitating further eliminations while carefully weighing the trade-off between cost and functionality. Following iterations of the flow, electrical impedance measurement emerged as a more cost-effective choice than optical methods. Regarding actuators, the team followed the proposed flow, ultimately selecting a peltier coupled with digital screen iterations.

\begin{figure}
    \centering
    \includegraphics[width=0.5\linewidth]{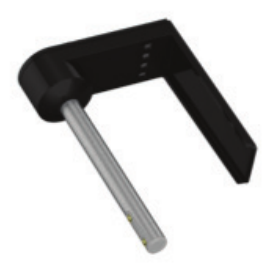}
    \caption{Custom electrical impedance measuring probe used in the yoghurt machine.}
    \label{fig:yoghurt_probe}
\end{figure}

\begin{figure}
    \centering
    \includegraphics[width=0.75\linewidth]{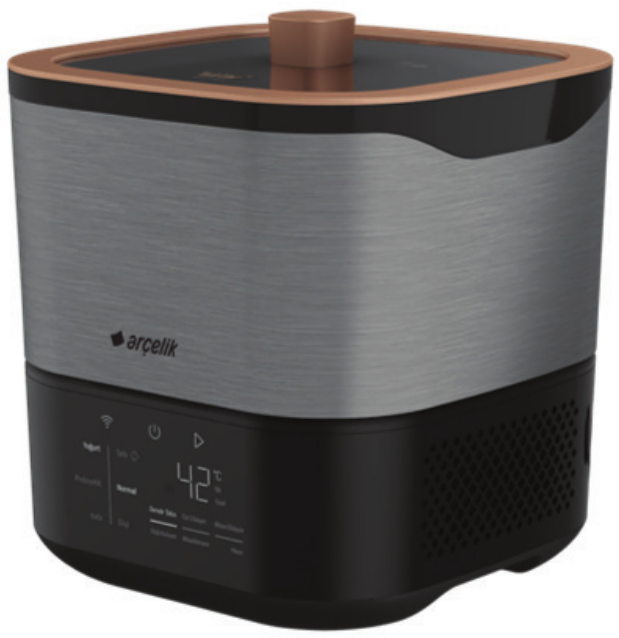}
    \caption{Final version of the yoghurt machine developed for mass production.}
    \label{fig:yoghurt_machine}
\end{figure}

\begin{figure}
    \centering
    \includegraphics[width=1\linewidth]{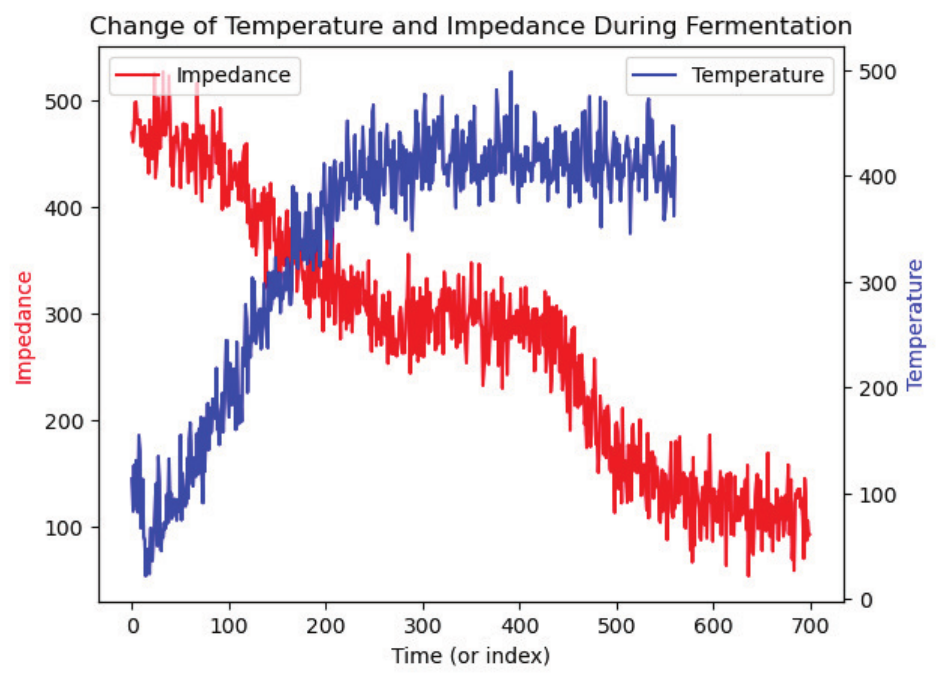}
    \caption{Change of impedance and temperature during fermentation.}
    \label{fig:yoghurt_impgraph}
\end{figure}

The current iteration of the yoghurt machine utilizes a custom, patented electrical impedance measuring probe (depicted in Figure \ref{fig:yoghurt_probe}), coupled with an NTC sensor to monitor the yoghurt's state continuously. In Figures \ref{fig:yoghurt_impgraph}, These data are visualized in an anonymous way with their axes names and values for confidentiality purposes. One can potentially argue that curves state a strong negative correlation, hence the inverted behaviour among electrical impedance and temperature. It is a known fact from physics that dairy liquids' impedance is inversely proportional to their temperature, meaning that higher temperatures result in a lower impedance under the same condition. However, the temperature settling time of milk is negligibly smaller than the whole fermentation duration, which ensures that temperature fluctuations do not affect the impedance measurement. The project successfully entered mass production and garnered positive user feedback, both during field tests and in real-world use. The final version of the device can be seen in Figure \ref{fig:yoghurt_machine}.

\subsubsection{Key Takeaways}
This project aimed to develop an autonomous yoghurt machine capable of sensing the acidity level and texture of yoghurt during the fermentation process. The machine is designed to be a compact household appliance, allowing users to customize the taste level of the end product before fermentation begins. A broad range of measurement and thermal actuation techniques were evaluated and tested, considering both performance and cost.

Approximately ninety people worked on this project from start to finish, including industrial designers, mechanical engineers, test and verification engineers, and domain managers. The proposed flow was utilized for the sensor and actuator selection for the project, and the whole project evolved from the idea to the product with the systematic help of the flow.

The project and iterations spanned four years from concept to serial production, and it is now selling thousands of units annually, meeting the needs of end users and allowing them to produce better yoghurt. Through iterations, the ideas were refined, the team avoided getting lost in the selection of sensors and actuators, and the iterative nature of the process allowed us to pivot when a particular sensor or actuator did not meet the requirements, ensuring the project and the team stayed on track toward the goals.

\subsection{Textile Type Detection Project}

\subsubsection{Project Overview}
The objective of this project is to develop a solution to the textile recognition problem in laundry. Sorting clothes by textile type and selecting the appropriate programs and settings for washing machines and tumble dryers can be difficult and error-prone for many individuals. To address this issue, the project team has formulated the following problem statement: \emph{"Sorting clothes by textile type and selecting the appropriate programs and settings during laundry is a challenge."}

To address the problem, the project team has identified two potential user interactions for recognizing textile types and sorting clothes accordingly. The first interaction involves using a handheld device to scan each textile and identify its type. Subsequently, the device would then present the user with various scenarios for loading the clothes into the machine and configuring the appropriate programs and settings. The second interaction entails using a sensing device attached to the machine to detect the type of textiles being added or removed from the machine. The device would then update the available programs and settings based on the identified textile types.

For detecting textile types, the project team has explored several textile properties that can be measured or detected. These properties include color, optical reflectivity, absorbance and emissivity, electrical conductivity and capacitance, and chemical composition (organic or inorganic). The team has also considered various sensor technologies and sensors capable of measuring these properties. These technologies include optical spectrum analyzers, optical spectrometers, optical intensity sensors, sensors detecting specific wavelengths, multi-spectral sensors, and LCR meters.

\subsubsection{Application of the Design Thinking Based Iterative Flow}

The table in Table \ref{tab:ssi_textile} displays the two iterations that have been used since the start of the project. The text that appears in \textit{italics} indicates that there have been no changes made from the previous iteration. It is important to note that revisiting and making changes on every step during each iteration is not mandatory. However, it is crucial to be meticulous because the flow is progressive and each step is built upon the previous one.

\begin{table}
    \centering
    \caption{Application iterations of the flow for Textile Type Detection Project}
    \includegraphics[width=1\linewidth]{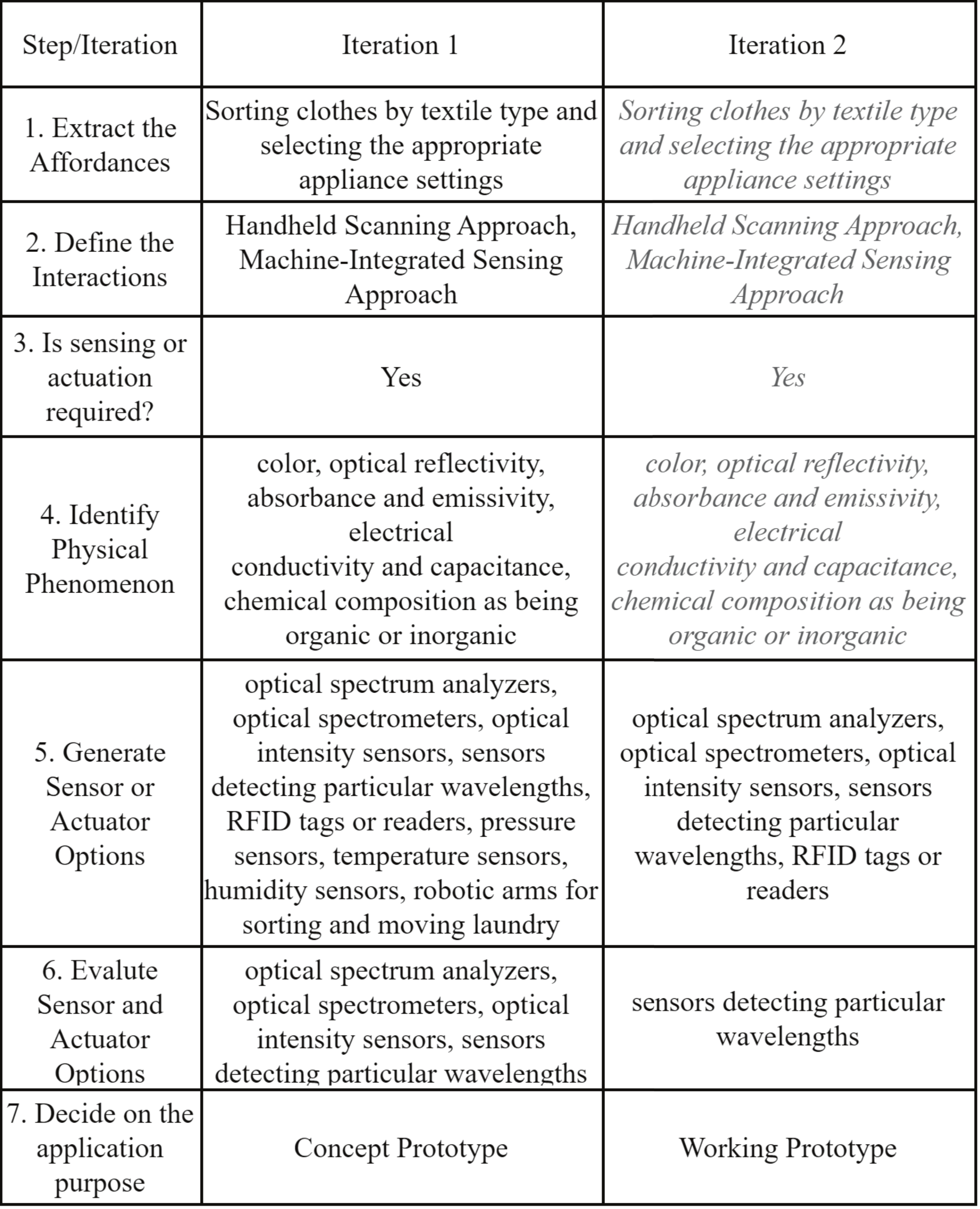}
    \textit{Italic} text represents no change compared to the previous iteration.
    \label{tab:ssi_textile}
\end{table}

\paragraph{\textbf{Extract the affordance(s)}} 
This project originated from the challenge of textile recognition in the laundry process. Many people find it difficult to accurately sort laundry before washing them in a washing machine or drying them in a tumble dryer. The affordance lies in the development of a device with the capability to identify the type of textile in each clothing item. A device capable of identifying the type of textile in each clothing item could simplify this task, making it easier to group similar textiles together and choose appropriate settings on washing machines or dryers. 
To clarify the starting point, we formulated a more specific problem statement (including the affordance): \emph{"Sorting clothes by textile type and selecting the appropriate appliance settings are challenging during the laundry process."}

\paragraph{\textbf{Define the interaction(s)}}
To address the identified problem of textile recognition in laundry, we need to define the interactions for the possible solutions. Here is the sequence of actions identified for our problem: Recognizing the type of the textiles, sorting the clothes accordingly, loading them into the machine (washer or dryer), and selecting the appropriate program and settings. Although various interaction sets could be designed, here are two examples:

\begin{itemize}
\item \textit{Handheld Scanning Approach:} The user scans each textile item with a handheld device, which identifies the type of the textile and display the result. Once all items are scanned, the device suggests various loading and program scenarios. For example, it might state "- load all the clothes into the machine and I will automatically set an appropriate programs and setting". Alternatively, it could suggest "- first, load the cotton-based clothes and I will set the appropriate program. Then add the synthetic items, and I will adjust the settings accordingly". 

\item \textit{Machine-Integrated Sensing Approach:} As the user places clothes into the washing machine or dryer, a built-in sensor identifies the types of textiles. The machine then dynamically updates its program and setting recommendations based on the combination of items loaded or removed. For example, the device might suggest "- you have added 15 cotton and black items so far, and the latest addition is synthetic and white. You could use a mixed program, but I recommend removing the last item.", or it might offer "- you have only loaded four synthetic items in different colors. The `Synthetics' program with the `Quick Wash' option would be suitable.". 
\end{itemize}

\paragraph{\textbf{Determine the need for sensing or actuating}}
Having defined the interactions, we must determine whether sensors or actuators are needed to implement them. While there may be non-sensor-based alternatives, such as reading textile types from clothing labels, our focus is on solutions involving sensors and/or actuators.  Therefore, it is necessary to identify the specific physical phenomenon to be sensed (or measured) and actuated. 

\paragraph{\textbf{Identify physical phenomenon}}
At this stage, the physical phenomena of interest might still be somewhat unclear. This is common and understandable, as available technology may not yet be sufficiently accurate or sensitive for the application. Identifying the appropriate phenomena typically requires comprehensive literature research and ideation, often involving specialists from relevant fields.

For our textile identification challenge, we considered the following properties of textiles to be measured or detected: color, optical reflectivity, absorbance and emissivity, electrical conductivity and capacitance, chemical composition as being organic or inorganic. Once these potential properties were identified, the next step involved exploring options of sensor technologies (and particular sensors). 

\paragraph{\textbf{Generate sensor and actuator options}}
After identifying the key physical properties to measure, we generated a list of potential technologies and specific devices for sensing and actuating. These include: optical spectrum analyzers, optical spectrometers, optical intensity sensors, sensors detecting particular wavelengths, multi-spectral sensors, LCR meters, infrared sensors, ultrasonic sensors, magnetic sensors, RFID tags or readers, pressure sensors, temperature sensors, humidity sensors, robotic arms for sorting and moving laundry, pneumatic actuators for opening and closing machine doors, electronic valves for controlling water and air flow in the machine Vibrating motors for shaking out debris from clothing, LED lights or display screens for indicating program and cycle status. As can be seen from Table \ref{tab:ssi_textile}, sensor options were narrowed down as they were evaluated in the first iteration. Combined with the fact that the application purpose shifted from concept prototype to working prototype, narrowing down the sensor options helped the project team to do more experiments with less sensors, yielding a more focused approach.

\paragraph{\textbf{Evaluate sensor and actuator options}}

As previously mentioned in related section, the evaluation of sensor and actuator options step is conducted using project-specific criteria. These criteria are unique to each project, and there are no fixed parameters or metrics, emphasizing the tailored approach to evaluation for each individual endeavor.

\begin{itemize}
\item{Operational feasibility:} Optical spectrum analyzers, optical spectrometers, optical intensity sensors, sensors detecting particular wavelengths, and multi-spectral sensors require complex and precise hardware and software configurations, which may demand trained personnel to operate. LCR meters, infrared sensors, ultrasonic sensors, magnetic sensors, RFID tags or readers, pressure sensors, temperature sensors, and humidity sensors are generally easier to set up and use, requiring less technical expertise. For the purpose of this project, AS7265X, which is an optical spectrometer made by AMS, was used to measure the distribution of the wavelengths reflected by the illuminated fabric. It can detect 18 different types of wavelengths, starting from 410nm to 940nm. The sensor package itself consists of three separate packages that work in conjunction. It can act as a slave device in an i2c bus, or can be controlled with UART, which allows flexibility in terms of communication.Robotic arms, pneumatic actuators, electronic valves, vibrating motors, LED lights, and display screens do require some level of technical expertise to effectively implement and operate.

\item{Financial feasibility:} Optical spectrum analyzers, optical spectrometers, and multi-spectral sensors are typically expensive, while sensors detecting particular wavelengths, LCR meters, infrared sensors, ultrasonic sensors, magnetic sensors, RFID tags or readers, pressure sensors, temperature sensors, humidity sensors, vibrating motors, and LED lights or display screens are more affordable. By the time that this project was live in 2018, the unit price for 1000 units of AS7265X was 2 USD, placing it in a competitive pricing point.

\item{Accuracy:} Optical spectrum analyzers, optical spectrometers, and multi-spectral sensors are capable of providing highly accurate measurements, while other sensors may offer lower levels of measurement accuracy. During textile scanning, the 3D printed housing of AS7265X acts as a shield to block environmental lighting, which act as noise. If the only light source is the onboard LED's with specific wavelengths near the sensor are present, the accuracy is maximized.

\item{Size feasibility:} Optical spectrum analyzers, optical spectrometers, and multi-spectral sensors are relatively large and may not be suitable for all applications, while other sensors are smaller and more compact.

\item{How easy is it to implement?:} Optical spectrum analyzers, optical spectrometers, optical intensity sensors, sensors detecting particular wavelengths, multi-spectral sensors, LCR meters, Infrared sensors, ultrasonic sensors, magnetic sensors, RFID tags or readers, pressure sensors, temperature sensors, and humidity sensors are all relatively easy to implement. Robotic arms, pneumatic actuators, electronic valves, vibrating motors, LED lights, and display screens may require more effort to integrate into a laundry sorting system.

\item{Resolution:} Optical spectrum analyzers, optical spectrometers, and multi-spectral sensors typically have high resolution capabilities, while other sensors may have lower resolution levels.

\item{Lifetime:} The lifetime of sensors and actuators can vary greatly depending on the specific technology and usage conditions. If an optical spectrometer is going to be used, it is imperative to not exceed the maximum brightness level that the sensor can handle. Otherwise, the lifespan that is indicated in their data sheets or user manuals will be void.

\item{Power requirements:} Some sensors and actuators may require more power than others, which can affect the overall power consumption of the system. For example, robotic arms and pneumatic actuators are generally more power-intensive, which could significantly increase the system's overall energy consumption. In contrast, temperature and humidity sensors typically consume less power. For reference, the LED driving circuit on the AS7265X board consumes 1.5W at 5V, if turned on continuously. Similarly, the measuring circuit consumes 1W at 3.3V, keeping the total power consumption around 2.5W, making the system compatible to work with USB adapters or batteries. Therefore, it is crucial to weigh the power requirements of each potential sensor or actuator should be considered when deciding on the most suitable components for an interactive laundry sorting system.

\item{Data transfer protocols:} The system should have a reliable and efficient method for data transfer between its sensors, actuators, and central processing unit. Since some sensors and actuators may use different communication protocols, additional hardware or software might be needed for seamless interaction. Therefore, the data transfer protocols of each component should be evaluated to ensure compatibility. For non-industrial use cases, UART and i2c are commonly available among sensor communication choices.

\item{Robustness to environmental factors:} Laundry sorting systems could be subject to fluctuating environmental conditions such as temperature, humidity, and dust. These factors can impact the accuracy and reliability of sensors and actuators. Therefore, it is important to select components that can withstand these environmental variables for long-term operation and reliability.

\item{Ease of maintenance and calibration:} Sensors and actuators may require regular maintenance and calibration to ensure they continue to function accurately. If their wavelength intervals are not exceeded by strong light sources at different wavelengths, optical spectrometers should require little to no calibration in their lifespan. The ease of these maintenance activities should be a factor when selecting components, as it affects both time and cost in the long run.

\item{Availability of support and replacement parts:} Sensors and actuators may occasionally fail and require replacement. Availability of replacement parts and technical support should be considered when selecting components, to ensure the system's sustained operation and minimize downtime.
\end{itemize}

\begin{figure}
    \centering
    \includegraphics[width=1\linewidth]{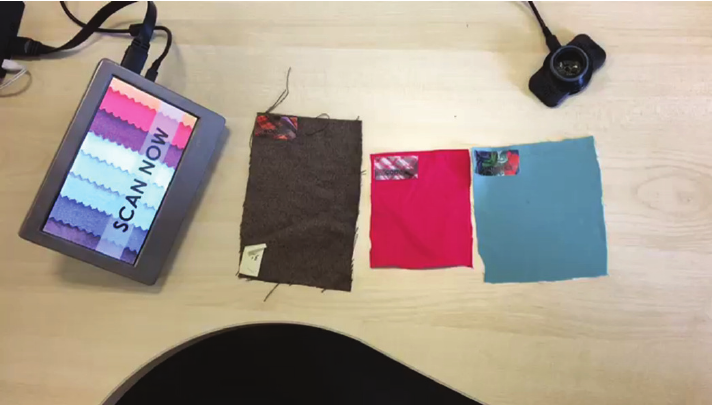}
    \caption{Prototype of the textile recognition system developed for the project.}
    \label{fig:textile-recog-prototype}
\end{figure}

\paragraph{\textbf{Decide on the application purpose}}
The goal of the Textile Type Detection Project was to develop a solution for accurately sorting clothes based on textile type and selecting appropriate laundry programs and settings. The project aimed to achieve mass production of the system for widespread adoption in domestic environments. For the textile recognition project, several prototypes targeting mass production were created to showcase the potential of the solution, one of which can be seen in Figure \ref{fig:textile-recog-prototype}.

Based on the thorough on-paper evaluation of sensor options, the project team selected multispectral optical sensors as the most suitable technology for textile type detection. These sensors demonstrated favorable performance across multiple evaluation criteria, including operational feasibility, accuracy, resolution, and robustness to environmental factors. Additionally, they offered the required level of accuracy and versatility needed to identify a wide range of textile types commonly found in laundry.

Though the project did not proceed to mass production stage, the selection of multispectral optical sensors reflects a strategic choice that would have supported the project's goal of offering an effective and efficient textile detection solution for domestic use.

\subsubsection{Key Takeaways}
This project's objective was to address the textile recognition challenge in laundry by accurately identifying the type of textiles and facilitating sorting clothes accordingly. To achieve this goal, the team has explored potential user interactions and assessed a wide range of sensor technologies. The evaluation process highlighted the importance of considering operational and financial feasibility, accuracy, size, ease of implementation, and other factors during the sensor technology evaluation phase. The selection of multispectral optical sensors, based on on-paper evaluation, demonstrates the project's strategic approach in pursuing a suitable solution. Collaboration with subject matter experts will be instrumental in further fine-tuning the chosen physical phenomenon to be sensed or actuated, thereby advancing the project's progress. The proposed flow allowed four engineers to collaborate with an industrial designer to temporarily finalize the project successfully. One of the bottlenecks in this project was the limited amount of physical space versus the desired functionalities. Since industrial designers may not have the same level of technical expertise as the engineers in the project, our flow helped the designers to focus on the affordances of each sensor, instead of their technical specifications. This way, industrial designers managed their timeline more efficiently, finishing the project in 2 years while it was planned to last for 2.5 years.

\subsection{3D Food Printer Project}

\subsubsection{Project Overview}

The objective of the 3D Food Printer project is to explore the integration of mobility into the domain of food printing, addressing the limitations of existing stationary 3D Food Printers (3DFPs). The research team identified that many commercial 3DFPs are adaptations of conventional 3D printing devices into the kitchen domain, resulting in stationary structures with limited printing capabilities. To advance the field, the team aims to introduce mobility to 3DFPs, exploring the potential of mobile food printing systems.

Inspired by swarm printing concepts \cite{oxman2014towards, robotsassemble, minibuilders, hunt20143d, 3dbot} that utilize synchronized small robots to construct large structures, the research team envisions a mobile 3DFP capable of printing edible materials on large surfaces. This novel approach could enhance dining experiences for children, provide personalized food options for patients in hospitals, and improve efficiency in professional kitchens. Furthermore, the mobility could facilitate the creation of larger food portions, expanding the technology's applicability.

The research question guiding this project is: \emph{"What if there was a printer that could print edible materials on large surfaces by moving freely?"}


\subsubsection{Application of the Design Thinking Based Iterative Flow}
The iterative flow for sensor and actuator selection regarding the 3D food printer project can be visualized in Table \ref{tab:ssi_3dfp123}.Below are the breakdown of each step as the flow is followed.  Italic text represents no change compared to the previous iteration.

\begin{table}
    \centering
    \caption{Application iterations of the flow for 3D food printer project}
    \includegraphics[width=1\linewidth]{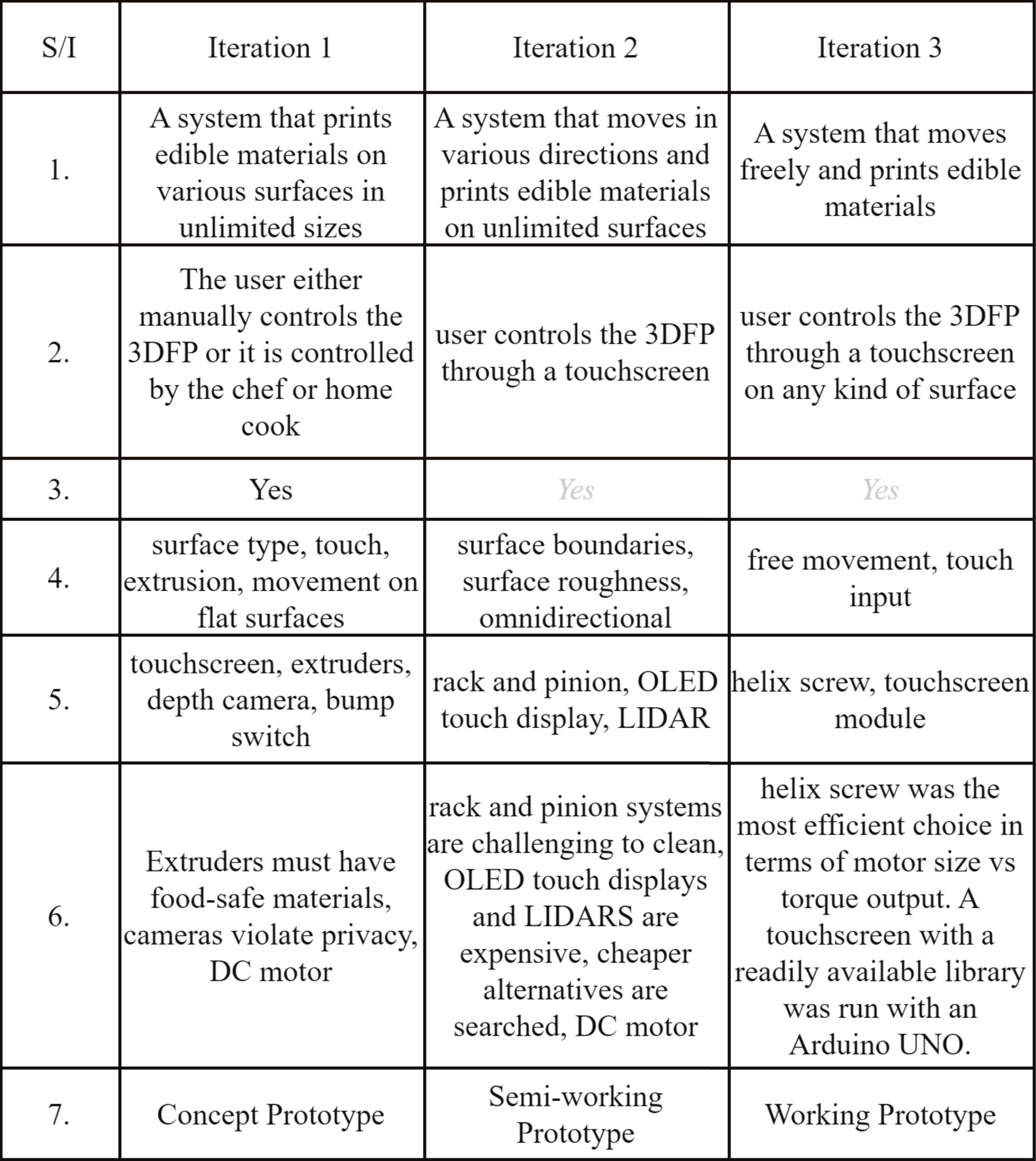}
    \textit{Italic} text represents no change compared to the previous iteration.
    \label{tab:ssi_3dfp123}
\end{table}

\paragraph{\textbf{Extract the Affordance(s)}}
The stationary structure of 3DFP devices limits their adaptability to the environment in which they are used. One of the major consequences of this rigidity is that it dictates the size of the printed item. Therefore, conventional 3DFPs have two main drawbacks: 1) a limited chassis and printing surface area that determine the maximum size of print-outs, and 2) unnecessary space occupation in the environment when they produce small print-outs or are not in use. To address these issues, we propose a novel 3DFP concept called the mobile 3D food printer (M3DFP).

\paragraph{\textbf{Define the interaction(s)}}

\begin{figure}
    \centering
    \includegraphics[width=1\linewidth]{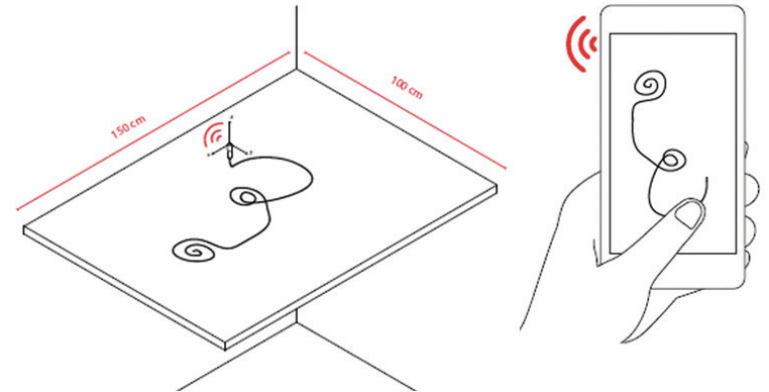}
    \caption{Mobile 3D food printer (M3DFP) principle drawing}
    \label{fig:m3dfp_principle}
\end{figure}

The challenges mentioned above motivated the research team to aim for the development of a working prototype of a mobile 3D printer (M3DFP) capable of printing edible puree materials on large surfaces. In this regard, we defined the main interactions as follows:
\begin{itemize}
\item Move freely on a predefined surface (initially 150x100cm, or preferably larger).
\item Print puree material on the predefined surface: M3DFP is expected to print food materials on surfaces of the users’ choice. It should be able to produce items in various sizes.
\item Be controlled by the user through a graphical user interface of any fidelity.
\end{itemize}
With these interactions defined, we created a drawing that essentially illustrates the working principle of M3DFP (Figure \ref{fig:m3dfp_principle}).

\paragraph{\textbf{Determine the need for sensing or actuating}}
The primary research question revolves around enabling the prototype to move freely on predefined large surfaces while allowing user control through a user interface. To identify the sensing and actuating requirements for this functional prototype, we divided the M3DFP system into four main components:

\begin{enumerate}
\item Effector: This component is primarily responsible for the movement of the printer. It implies the need for actuator motors and some form of wheels. Additionally, this component highlights the necessity for sensors to assist in navigating the printer on the printing surface.
\item Arm: This component positions the extruder away from the platform to prevent overlaps during the printing process. The arm is designed to be stable, eliminating the need for actuating or sensing.
\item Extruder: To extrude the material onto the surface, there is a requirement for actuators that move the screw inside the syringe filled with edible materials.
\item User Interface: For the current prototype, the research team did not specify any fidelity requirements, as this would be a research prototype \cite{olson2014ways}. Consequently, the initial interface will consist of a smartphone screen.
\end{enumerate}

\paragraph{\textbf{Identify physical phenomenon}}
The operation of the machine raises fundamental questions such as "how will the device determine where to print?", "what is the size of the printing space?", and "at what rate will the M3DFP extrude materials?". These inquiries highlight the importance of measuring both distance from a reference point and material flow.

To address these physical phenomena, we consider the following:

\textbf{Distance Measurement:} Determining the machine's distance from the edges of the predetermined surface (initially 150x100 cm) is crucial.

\textbf{Material Flow Control:} It's essential to regulate the amount of edible material extruded from the syringe.

Additionally, to mitigate potential rheological challenges associated with the edible material, we've opted to begin with a readily manageable substance. Whipped cream, known for its controllable pressure and viscosity, was selected. This choice not only simplifies preparation but also aligns with the project's objectives.

This approach ensures that the chosen materials are conducive to the successful operation of the mobile 3D food printer.

\paragraph{\textbf{Generate sensor and actuator options}}
The options below have been devised to enable the M3DFP to interact effectively with the environment, ensuring accurate and controlled printing on large surfaces based on identified physical phenomena. Please note that not all of the following options are listed in Table \ref{tab:ssi_3dfp123} for simplicity's sake. It is imperative to observe that both sensor and actuator options were narrowed down as iterations took place. Researchers were more keen on giving details about the options, rather than coming up with other options as iterations were taken. This nature of an iterative process helped the researchers to converge towards possible solutions.

\textbf{Sensor Options:}

\begin{itemize}
    \item LIDAR Sensors: Accurate distance measurement for precise positioning.
    \item Distance Sensors: Provide exact distance measurements for surface proximity.
    \item Flow Sensors: Monitor extrusion rate for controlled printing.
    \item Infrared Sensors: Enable proximity detection for component interaction.
    \item Force Sensors: Measure extrusion pressure for consistent flow.
    \item Vision Systems: Offer visual feedback for navigation and control.
    \item Load Cells: Monitor syringe weight for material control.
\end{itemize}

\textbf{Actuator Options:}

\begin{itemize}
    \item Stepper Motors: Drive omni-wheels for surface movement.
    \item DC Motors: Control extruder screw for material flow.
    \item Motorized Syringe Plunger: Regulate pressure and flow.
    \item User Interface (Smartphone Screen): Input commands and monitor process.
    \item Pneumatic Actuators: Ensure precise control over syringe pressure.
    \item Linear Actuators: Adjust position relative to the printing surface.
    \item Piezoelectric Actuators: Enable fine-scale adjustments.
    \item Voice Coil Actuators: Provide rapid and precise motion.
\end{itemize}

\paragraph{\textbf{Evaluate sensor and actuator options}}

\textbf{Sensor Options:}

\textbf{LIDAR Sensors:} LIDAR sensors offer high precision in distance measurement, making them suitable for accurate positioning of the M3DFP. Their ability to create detailed 3D maps of the surroundings can enhance the printer's navigation capabilities.

\textbf{Distance Sensors:} These sensors provide reliable distance measurements, ensuring that the M3DFP maintains consistent proximity to the printing surface. They are essential for preventing collisions or variations in printing height.

\textbf{Flow Sensors:} Flow sensors play a crucial role in regulating the extrusion process. By monitoring the rate at which edible material is dispensed, they contribute to controlled and precise printing. This ensures that the printed patterns and structures maintain their intended integrity.

\textbf{Infrared Sensors:} Infrared sensors offer proximity detection capabilities, allowing the M3DFP to sense the distance between its components and the printing surface. They can be employed to provide additional safety measures and improve overall accuracy during printing.

\textbf{Force Sensors:} Force sensors are instrumental in measuring the pressure applied during the extrusion process. By ensuring a consistent force, they contribute to uniform material flow, which is essential for achieving high-quality prints with consistent structural integrity.

\textbf{Vision Systems:} Vision systems equipped with cameras and advanced algorithms provide visual feedback that aids in navigation, object recognition, and precision control during printing. They enhance the M3DFP's ability to adapt to dynamic environments and adjust its printing parameters accordingly.

\textbf{Load Cells:} Load cells are valuable for monitoring the weight of the syringe, enabling precise control over material usage. By tracking the remaining material, the M3DFP can optimize its printing process and avoid interruptions due to insufficient supplies.

\textbf{Actuator Options:}

\textbf{Stepper Motors:} Stepper motors \cite{stepmotor} offer precise control over movement, making them suitable for driving the omni-wheels. They enable the M3DFP to navigate smoothly across the printing surface, adjusting its position as needed.


\textbf{DC Motors:} DC motors play a crucial role in controlling the extruder's screw mechanism. By regulating the flow of edible material, they ensure consistent and accurate printing. The speed and direction of the screw can be adjusted to accommodate various printing requirements.

\textbf{Omni Wheels} These wheels allow movement in various axes, thus making the prototype moving freely, in any direction. They allow the platform to move infinitely possible directions without maneuvering.

\textbf{Motorized Syringe Plunger:} This actuator option provides a means to regulate the pressure and flow of material from the syringe. By controlling the plunger's movement, the M3DFP can fine-tune the extrusion process, resulting in precise and controlled printing outcomes.


\textbf{User Interface (Smartphone Screen):} A graphical user interface on a smartphone screen serves as the primary means of interaction between the user and the M3DFP. It allows users to input commands, customize printing parameters, and monitor the printing process in real-time. The user-friendly interface enhances the overall usability of the system.

\textbf{Pneumatic Actuators:} Pneumatic actuators provide precise control over syringe pressure, allowing for adjustments during the printing process. By modulating the air pressure, the M3DFP can achieve consistent extrusion rates and adapt to different material viscosities.

\textbf{Linear Actuators:} Linear actuators enable controlled vertical movement, allowing the M3DFP to adjust its position relative to the printing surface. This capability is valuable for maintaining a consistent distance from the surface, especially when printing on irregular or uneven substrates.

\textbf{Piezoelectric Actuators:} Piezoelectric actuators offer fine-scale adjustments, allowing for precise control over the extrusion process. Their ability to produce small, controlled movements is advantageous for achieving intricate details in the printed patterns.

\textbf{Voice Coil Actuators:} Voice coil actuators provide rapid and precise linear motion, which can enhance the printing accuracy of the M3DFP. Their responsiveness and high-speed operation make them suitable for dynamic printing scenarios that require quick adjustments.

Each of the mentioned sensor and actuator options brings specific strengths and considerations to the design of the M3DFP. The evaluation of these options is essential for selecting the most suitable components that align with the project's objectives and requirements.

\paragraph{\textbf{Decide on the application purpose}}

As the current prototype will be a research prototype for a Research Through Design project \cite{olson2014ways}, we initially considered cost and time efficiency while making the final decisions. 

\begin{figure}
    \centering    
    \includegraphics[width=1\linewidth]{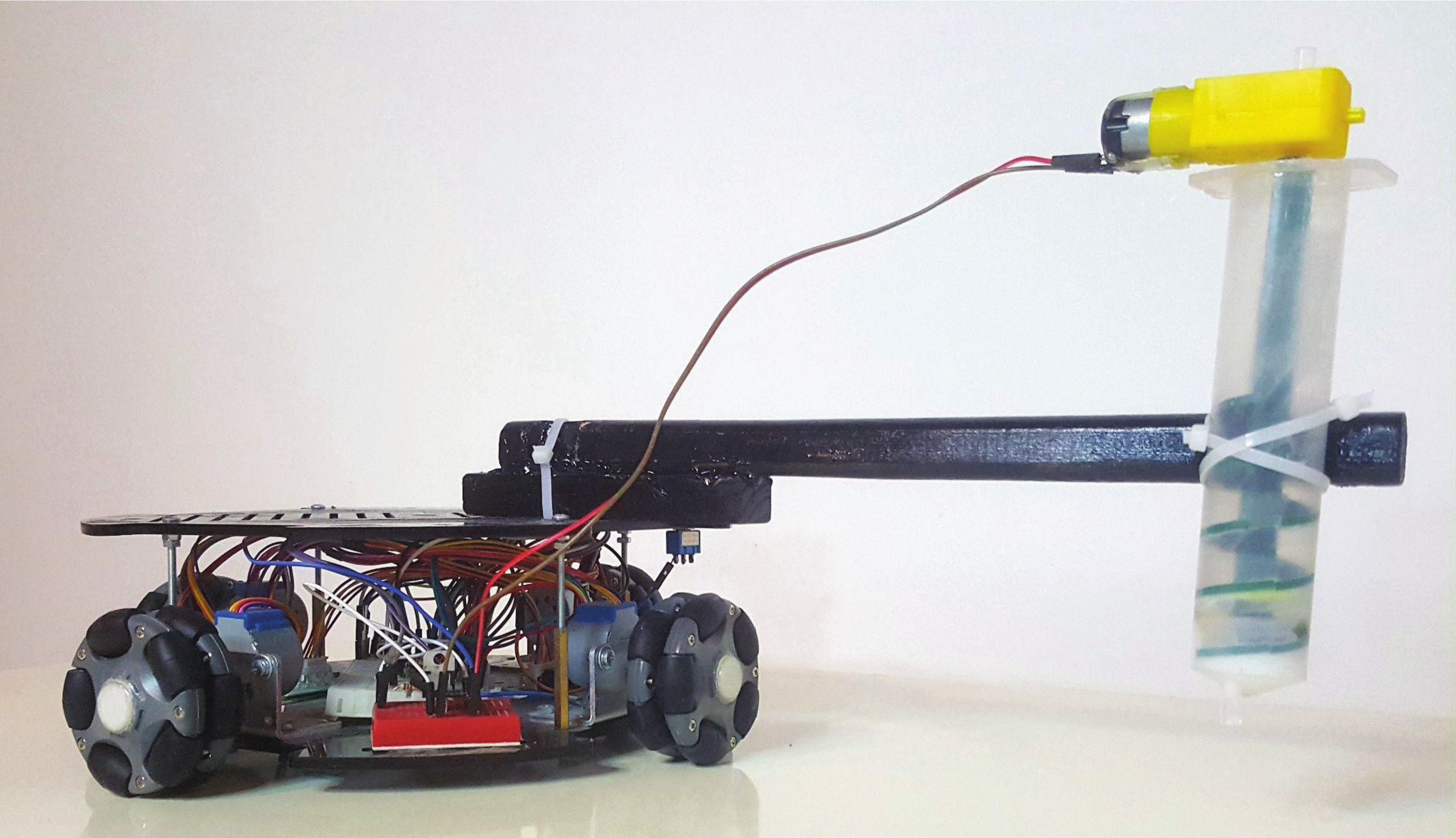}
    \caption{Final research prototype of M3DFP project.}
    \label{fig:m3dfp-prot}
\end{figure}

As a result of our ideation in iteration 1, we decided to employ a preexisting mobile omni-wheel robot and implemented our remaining parts on the device \cite{omnirobot}. We pursued this option as it was the most time efficient one for obtaining initial information from the prototype. With this device selection, we eliminated the need for building omni-wheels and the software to operate them from scratch. This helped the team focus on developing the extrusion mechanism and the user interface.
For the extrusion mechanism, we implemented a 3D printed screw inside a 50cc syringe attached it on a fixed arm. We designed the screw from scratch and in three iterations we reached the helix screw \cite{carley1953basic}. The extruder consists of a 50cc syringe used as a tube (i.e., food container) and a 3D printed polylactic acid (PLA) screw inside the syringe driven by a DC motor. The motor on the screw is powered by an external power source located on the body. The platform, arm, and extruder were attached to the omniwheel body using plastic clamps (Figure \ref{fig:m3dfp-prot}).


\subsubsection{Key Takeaways}

Visualizing the basic working principle of the device enabled both the research team and the engineers to communicate effectively, fostering a better understanding of potential ways to build the prototype. In this project, we learned that making the best decision does not always require building components from scratch. Instead, we explored various methods to integrate pre-existing technological components into our prototype building process. This approach was highly favored by designers because implementing the sensor and actuator selection flow in an iterative nature while building the M3DFP prototype allowed them to utilize skills from their own workflows. With the assistance of the proposed flow, engineers and the research team were able to present a concept prototype that would act as a baseline for the motivation, which was to develop a research prototype that would help discover possible food futures.

\subsection{Gesture Recognition Project}

\subsubsection{Project Overview}
The objective of this project is to expand the potential use cases of an existing two-hand gesture set that as been previously explored within a single domain \cite{hsos}. To address this limitation, the researchers formulated the following problem statement: \emph{"The validity of the-two hand gesture set has been examined only within a single domain, leaving other potential use cases undefined or uncertain."} 

To address this challenge, the researcher devised a strategy based on the Research-Through-Design methodology \cite{10.1145/1240624.1240704}. The suitability of this approach became evident due to its incorporation of iterative processes during both technical and design development phases.

\begin{figure}
    \centering
    \includegraphics[width=0.49\linewidth]{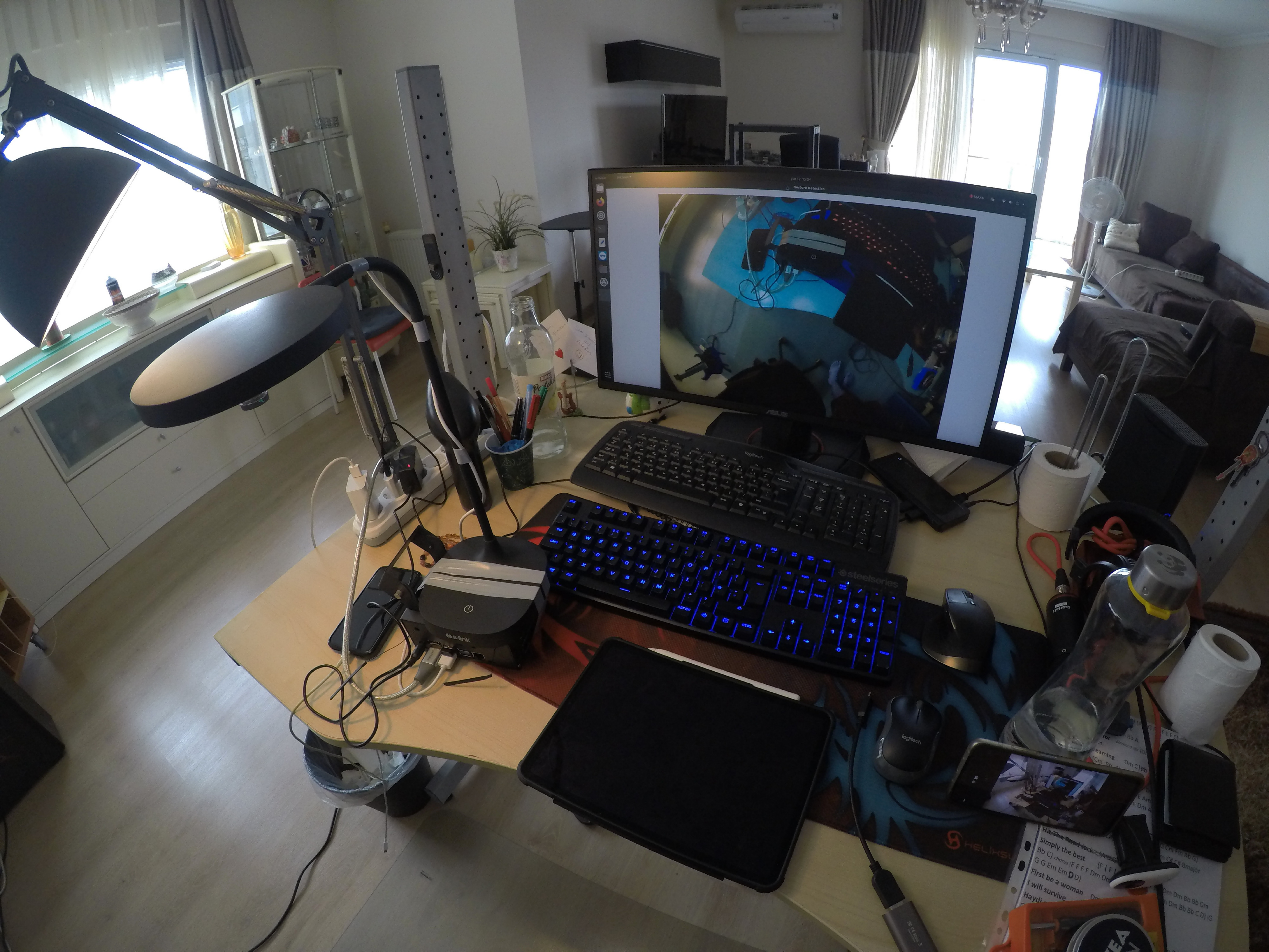}
    \includegraphics[width=0.49\linewidth]{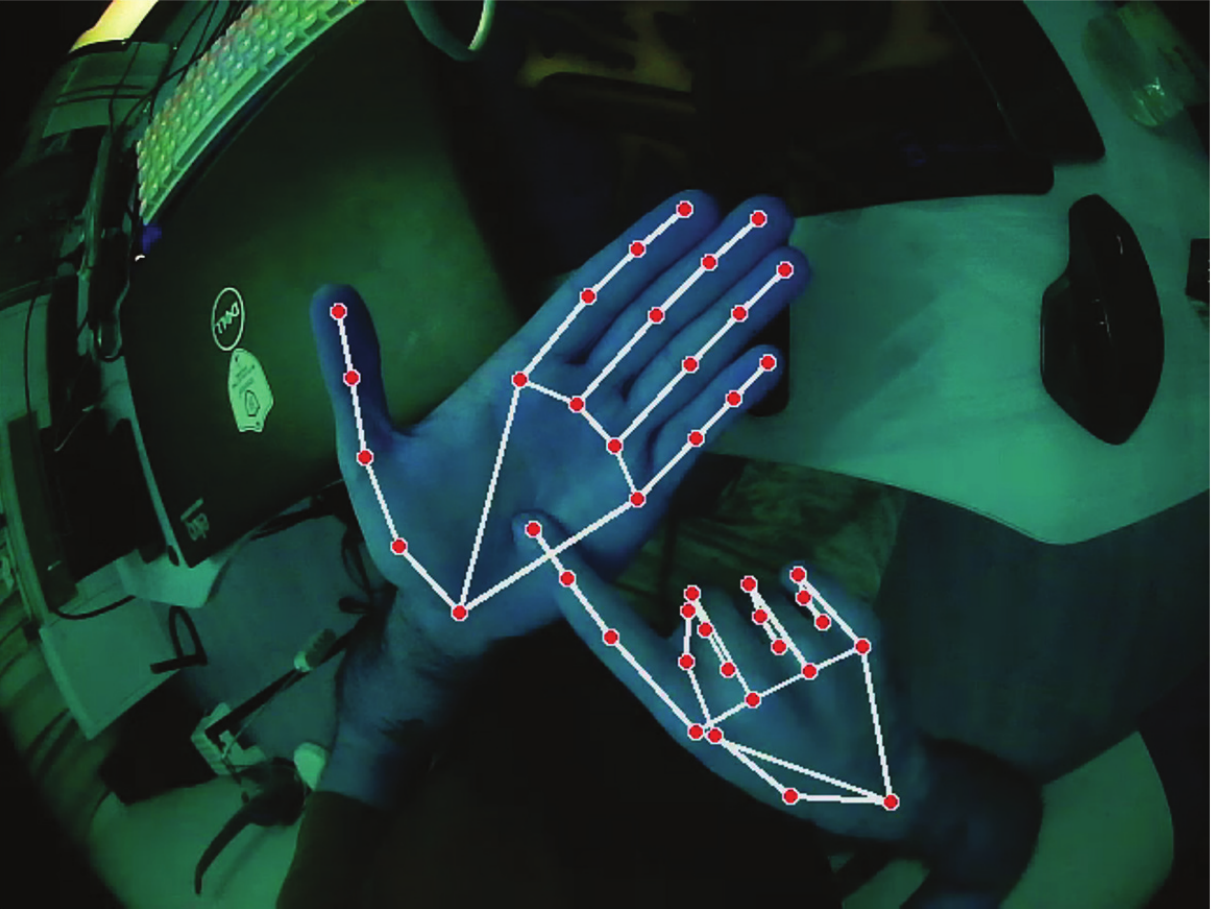}
    \caption{Two hand gesture recognition prototype}
    \label{fig:gesture_list}
\end{figure}

The technical aspect of the project commenced with a comprehensive literature review of the possible gesture detection methods. These methods were categorized based on their physical properties, encompassing aspects such as vision, acoustics or optics. Evaluation criteria included functionality and the abundance of online documentation for each method. Through a natural process of elimination inherent in the proposed flow, the researchers identified a single sensing method. 

\subsubsection{Application of the Design Thinking Based Iterative Flow}

The iterative flow for sensor and actuator selection regarding the gesture recognition project can be summarized in Table \ref{tab:ssi_ges123}. This table is an outline of the steps that the researcher followed while conducting his study.  Italic text represents no change compared to the previous iteration.

\begin{table}
    \centering
    \caption{Application iterations of the flow for Gesture Recognition Project}
    \includegraphics[width=1\linewidth]{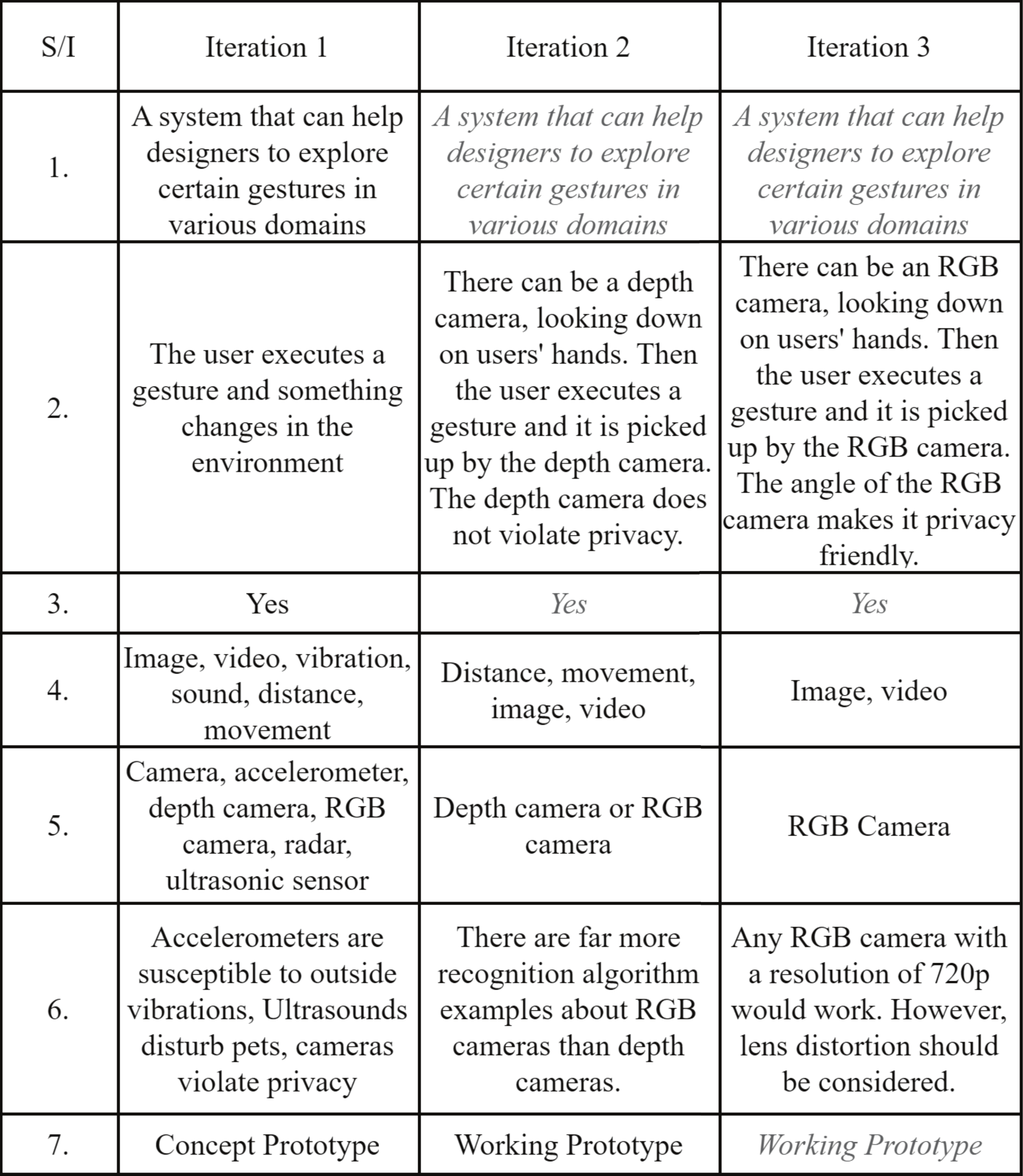}
    \textit{Italic} text represents no change compared to the previous iteration.
    \label{tab:ssi_ges123}
\end{table}

\paragraph{\textbf{Extract the affordance(s)}} 
The main motivation for this research is to explore further possibilities and use cases for a two hand gesture set that is specific to a domain. Similarly, the research also aims to help other gesture enthusiasts to explore other domains with certain gesture sets. The outcome of the research should afford to guide other researchers that want to further extend the use cases of their desired gesture set. 

\paragraph{\textbf{Define the interaction(s)}}

Multiple interaction possibilities regarding the system-to-be-developed were laid out, including:
\begin{itemize}
    \item Worn devices 
    \item Table top devices
    \item Portable devices
\end{itemize}

A worn device enables mobility, but sacrifices dependence in the form of battery charging. A table top device lacks in mobility, but is superior in computing power due to its larger form factor. Finally, a portable device comes with the drawback of battery dependence.

\paragraph{\textbf{Determine the need for sensing or actuating}}
A working prototype that is capable of classifying gestures and giving feedback is desired. As a result of this, the project needs sensors and actuators.

\paragraph{\textbf{Identify physical phenomenon}}
Two hand gestures translate into hand movements in a broader view. Instead of asking \emph{"How to detect gestures"}, asking \emph{"How to detect hand movements"} yields more generic results. A minimalist research indicates hand movements can be tracked with capacitive measurements \cite{wong2021multi}. Similarly, they can be tracked with visional methods such as cameras and light. 

\paragraph{\textbf{Generate sensor and actuator options}}
As we now know what to measure, we can lean towards online research with the following question(s): \emph{"How can we detect gestures"}, \emph{"How can we detect change in capacitance for hand movements"}. The results comprise of but are not limited to radars \cite{lien2016soli}, depth cameras \cite{ren2011depth} and RGB cameras \cite{van2011combining}.

\paragraph{\textbf{Evaluate sensor and actuator options}}

In this phase of the proposed flow, each sensor type that was listed previously was evaluated one by one. Their strengths and weaknesses were laid out specifically for gesture detection and their relevance to the subject was identified. An isolated research among all of the options yielded the following results.

\begin{itemize}
    \item \textbf{Radar:} 
    Radars work by emitting electromagnetic waves and expecting to receive them as they reflect from obstacles. They are robust against harsh climates and can work with no light present. However, a deeper research shows advanced signal processing techniques are required to set a custom radar system up for gesture detection. 
    
    \item \textbf{Depth camera:} 
    Depth cameras use IR (infra-red) light and receivers to calculate the distance of an object. Research shows that they are one of the strongest choices when it comes to gesture sensing. Combined with the fact that their ease of setup, depth cameras were considered as a suitable solution for the case of two hand gesture recognition. 
    
    \item \textbf{RGB camera:} Almost everyone with a smartphone carry an RGB camera with them in our current world. They can capture a frame consisting of color values for each pixel. For the world of gesture sensing, this means hands can be tracked accurately. Combined with the fact that they come in a variety of resolutions and specifications, they were considered as a strong alternative for gesture sensing. 
\end{itemize}

\paragraph{\textbf{Decide on the application purpose}}
We want to grasp an idea of how a working prototype would look like and function if it was put together. This project does not have any mass production goals. However, the findings from this study will be used in other gestural research.

\begin{figure}
    \centering    
    \includegraphics[width=0.9\linewidth]{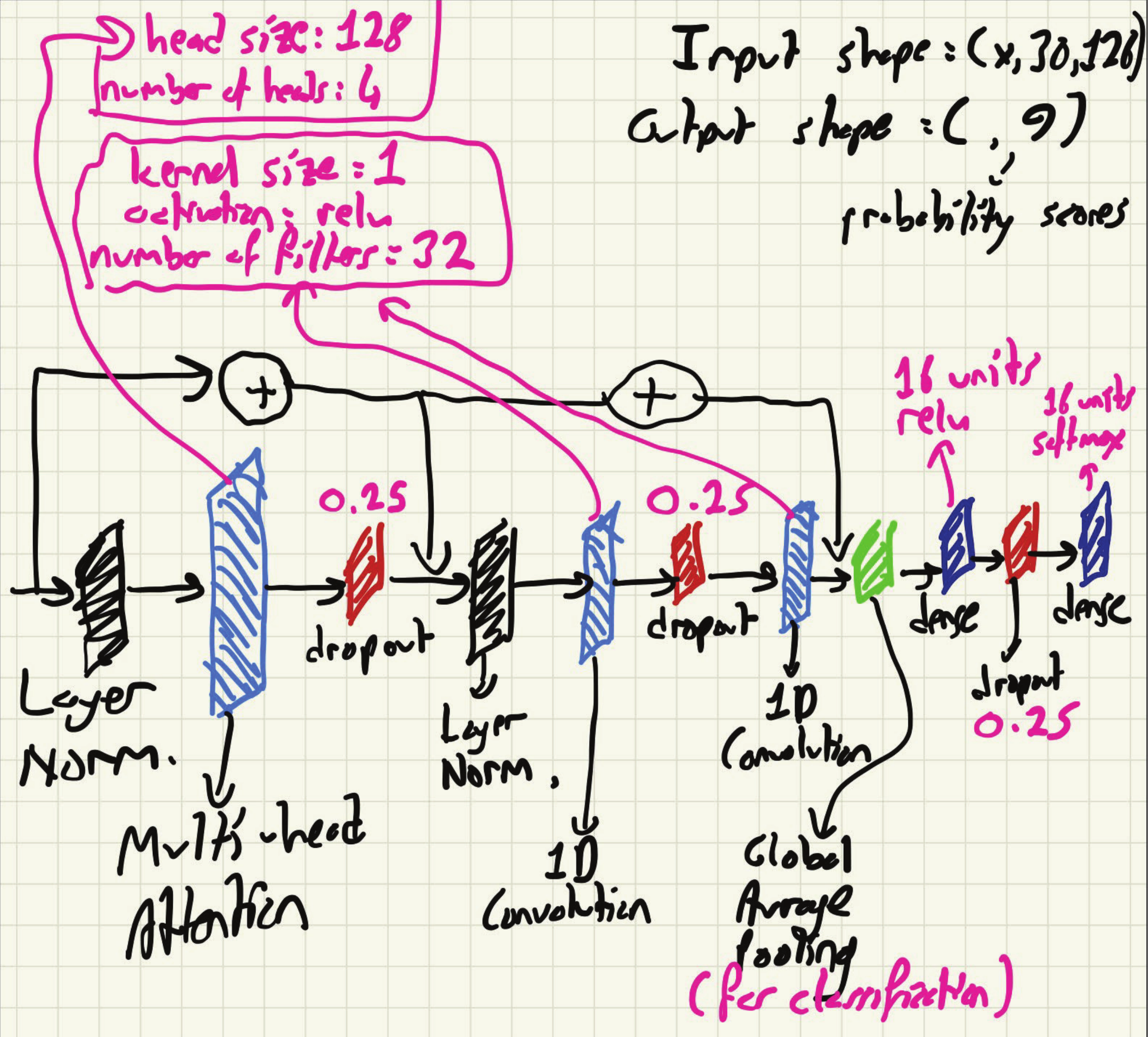}
    \caption{The on-paper prototype of the proposed transformer network to be used for gesture classification\cite{transformer}}
    \label{fig:gesture_network}
\end{figure}

As a result of following the proposed flow, the sensing methodology was isolated to RGB cameras. Similarly, while selecting among possible classifier networks, the output of the sensor and its data structure was considered. As a result, a transformer network based on multi-head attention was used \cite{transformer}. An on-paper sketch of the aforementioned network can be seen in \ref{fig:gesture_network}, which was used as a starting point to construct the neural network for gesture classification. The research agenda is progressing as planned. The study is halfway complete and user studies are being planned to try out the prototype in the wild. 

\subsubsection{Key Takeaways}
In this project, the researcher aimed to explore further the possible use cases of a two hand gesture set. A review of the already existing gesture detection methods was necessary, which is then fed to the technical development phase. Expected contributions include a working prototype for better immersion in user studies and design insights regarding exploration of gesture sets in different domains In the current state of the project, the researcher eliminated the possible sensing options to an RGB camera in a timeline of approximately 2 months. A major challenge during this research was the abundance of literature in terms of gesture recognition. Having such a diverse array of sensing methods made it difficult to filter them appropriately. However, the utilization of the selection flow allowed iterations. This is why the researcher could systematically go back and forth between steps, validating the choices he/she had made. 


\section{Discussion}
Below, we assess the results achieved by implementing our \emph{Design Thinking Based Iterative Sensor and Actuator Selection Flow} in the previously outlined projects. Additionally, we explore the connections between these outcomes and the existing body of previous research.

\begin{itemize}
    \item \textbf{Autonomous yoghurt machine:} Employing the proposed flow, the yoghurt maker project successfully made it into mass production, finding new users daily. An extensive literature research was performed as a part of \emph{Generate sensor and actuator options} step during the iterations. As the project goals were clarified, more and more of the listed sensors were eliminated as a part of the flow. Multiple iterations were done regarding the \emph{possible sensors} list, resulting in a refined array. While applying our proposed flow, our designerly perspective prevented us from tunnel visioning to a specific sensor type. \textbf{The project is currently in mass production, being validated by hundreds of field test users.}
    \item \textbf{Textile type detection:} Following the proposed flow led to successfully completing the project's feasibility study. Various experiments for different use cases were carried out, ensuring the validity of our sensor selection procedure both in technical and user-centric aspects. Notably, our flow focuses primarily on identifying appropriate sensor technologies, leaving subsequent technical details to specialized teams. Compared to a more technically-oriented pre-research approach, our designerly flow expedited the project timeline, allowing us to meet all deadlines. \textbf{The project has been archived as completed}, and all the know-how was reported.
    \item \textbf{3D Food printer:} As a result of applying our proposed method, a list of potential actuator and sensor options were listed for a mobile 3D food printer. The topic needed such an approach because it differs from regular 3D printers in many aspects and an iterative, designerly approach was necessary to develop suitable options. \textbf{The study is part of a research through design project, which will be finalized in the near future.} 
    \item \textbf{Gesture recognition:} The technical phase of the study is complete, thanks to a rapid sensor selection process which is proposed in this article. The research aims to explore further use cases for a specific, two-hand gesture set by means of working prototypes. The project was kicked off from a highly technical point of view, which also meant the sensor selection process was initially technical. This caused the researcher to be \emph{lost} in datasheets, resolutions and numerical values while completely unseeing the affordance of the problem. When the sensor selection process was restarted with our proposed method, possibilities were laid out, evaluated, and isolated in a matter of 2 weeks, which allowed the technical implementation phase to start. \textbf{This research is part of a doctoral research, which is proceeding parallel with the planned agenda. }
\end{itemize}

Through iterations, the ideas became more refined, ensuring the teams didn't lose track while selecting sensors and actuators, and the iterative nature of the process allowed them to pivot when a particular sensor didn't meet the requirements or expectations, ensuring the stayed on track toward ultimate project goals.

As a common theme across all projects:
\begin{itemize}
\item The application of the proposed flow resulted in a time-saving boost, enabling researchers and developers to surpass their initial timelines and alleviate time constraints.
\item Each project reached a significant milestone by adopting the proposed flow.
\item The teams expanded their repertoire of sensors and actuators, equipping them for future endeavors, including different projects or research pursuits.
\item Following an iterative, designerly approach emphasized the importance of the process itself over the final outcome.
\end{itemize}

We have substantiated our previous claims by testing our proposed flow in four different projects with completely different goals. As a result, we have validated the relevance of our solution towards sensor and actuator selection for interactive system design. 


Through our studies, we have reached the conclusion that our proposed flow for selecting sensor and actuator technologies can be instrumental in addressing interactive system design challenges. These challenges encompass a wide range of domains, and our flow can assist in making informed decisions about sensors and actuators, whether or not a functional prototype is required. Our study is applicable across various industries including food, beverages, textiles, and gestural interactions, making it highly robust. We are confident that as more researchers adopt our proposed flow, its contributions to the field will gain further validation.

It's important to note that our current study focuses on the elimination process of sensor and actuator options. Future research endeavors may involve integrating subsequent steps into our designerly approach. Additionally, refining the \emph{"Evaluate sensor and actuator options"} step to include numerical values and technical terminology could enhance its effectiveness.

\section{Conclusion}

The central research question propelling this study was, \textit{"How might teams of collaborating designers and engineers efficiently select sensors or actuators for their interactive system design and prototyping projects?"} We have proposed an actionable and tangible flow in response based on our previous experiences in interaction design projects. Following an approach parallel to Design Thinking, we have answered our research question with an easy-to-follow model that teams consisting of designers and engineers can adapt to their projects. This flow navigates the complex and often convoluted process of sensor and actuator selection, mitigating challenges that may arise from interdisciplinary collaborations. As demonstrated in this article, using this workflow in interaction design projects, teams can decrease the complexity, time, and costs while increasing the fluency of sensor and actuator selection processes.

Our article showcases its versatility and effectiveness across different interactive design scenarios through the lens of four real-life case studies. These case studies serve not only as empirical proof but also as a practical example for the audience, demonstrating the flow's applicability in varying contexts and project scales. After presenting our case studies and demonstrating the effectiveness of our proposed flow, we conclusively address our research question, showcasing how our approach significantly streamlines the sensor and actuator selection process for interactive system design.

In terms of its broader impact on academia, this article serves as a foundational piece for future research. The proposed flow and considerations can act as a valuable template for similar initiatives, enhancing our understanding of efficient and effective decision-making processes in design and engineering. The reliability of our proposed flow is evidenced by its successful application across various real-world projects, each with distinct goals and requirements. The iterative nature of our approach ensures adaptability and reliability in diverse design and engineering contexts.

Beyond its academic relevance, this article also contributes materially to industrial practices. By offering a structured methodology that is immediately applicable in real-world settings, we not only help project teams save valuable time but also enhance the quality of the end product. The provision of a road map that clearly demarcates the journey from problem identification to sensor or actuator implementation allows for more effective resource allocation and execution of project goals. Moving forward, we envision further refinement and adaptation of our flow to cater to an even wider range of interactive system design projects. By providing a foundational framework, we hope to inspire future research that advances the field of sensor and actuator selection.

In summary, this article goes beyond mere theoretical considerations to offer practical insights and tools that help answer our core research question comprehensively. While our approach has its limitations, its contributions to both the academic and industrial landscapes are invaluable. As our method is increasingly adopted and adapted, we are optimistic that it will serve as a cornerstone for further research, offering a more mature and generalized approach to the complex dynamics of interdisciplinary collaborations. Our work lays a foundation for a more streamlined and user-centered process in selecting sensors and actuators, significantly benefiting the practice of interactive system design.

\bibliography{dc-sa}{}
\bibliographystyle{IEEEtran}

\begin{IEEEbiography}[{\includegraphics[width=1in,height=1.25in,clip,keepaspectratio]{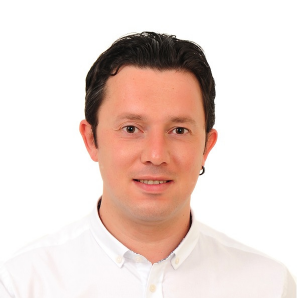}}]
{İhsan Ozan Yıldırım} received his B.Sc. degree in Electronics and Communications Engineering from Yıldız Technical University in Istanbul, Türkiye, in June 2010. He obtained his M.Sc. degree from the Electrical and Electronics Engineering Department at Koç University,Istanbul, Türkiye, in January 2013. Currently, he is actively pursuing a Ph.D. degree in Design, Technology, and Society at Koç University - Arçelik Research Center for Creative Industries, İstanbul, Türkiye.

Presently serving as a Senior Lead Engineer in the Sensor Technologies Directorate at Arçelik A.Ş. R\&D Center, Çayırova, İstanbul, Türkiye. His research focus revolves around smart device prototyping, sensor system development, air quality sensing, and data science applications related to sensor technologies in consumer electronics.
\end{IEEEbiography}

\begin{IEEEbiography}[{\includegraphics[width=1in,height=1.25in,clip,keepaspectratio]{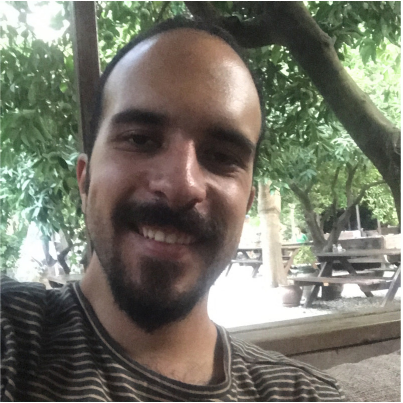}}]{Ege Keskin} received his Bachelor's and Master's degrees in Mechatronics Engineering from Bahçeşehir University, Istanbul, Türkiye. He is currently pursuing a Ph.D. degree in Design, Technology, and Society at Koç University - Arçelik Research Center for Creative Industries, Istanbul, Türkiye.

Ege Keskin also currently serves as a Senior Specialist Engineer in the Sensor Technologies Directorate at Arçelik A.Ş. R\&D Center, Çayırova, Istanbul, Türkiye. His primary research interests lie in the development of sensor technologies for domestic appliances. Moreover, he is actively engaged in exploring the applications of deep learning techniques, particularly in the field of gesture recognition systems.
\end{IEEEbiography}

\begin{IEEEbiography}[{\includegraphics[width=1in,height=1.25in,clip,keepaspectratio]{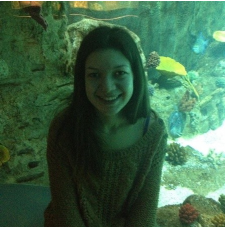}}]{Yağmur Kocaman} is a Ph.D. student currently pursuing her degree in Koc¸ University - Design, Technology and Society program. Originally trained as an industrial designer, she currently researches about designing
novel food making technologies for future homes. Her research
interests are human-centered design and human-food interaction.
\end{IEEEbiography}

\begin{IEEEbiography}[{\includegraphics[width=1in,height=1.25in,clip,keepaspectratio]{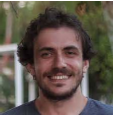}}]{Murat Kuscu} received his PhD degrees in Engineering from University of Cambridge, UK, in 2020, and in Electrical and Electronics Engineering from Koc University, Turkey, in 2017. He is currently an Assistant Professor at the Department of Electrical and Electronics Engineering, Koc University. His research interests include Internet of Bio-Nano Things, nanomaterials, biosensors, and microfluidics.
\end{IEEEbiography}

\begin{IEEEbiography}[{\includegraphics[width=1in,height=1.25in,clip,keepaspectratio]{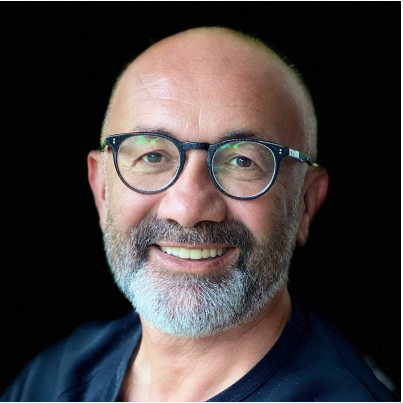}}]{Oğuzhan Özcan} is professor of interactive media design at Koç University‘s Department of Media and Visual Arts, and the director of the Koç University – Arçelik Research Center for Creative Industries.

Özcan has degrees in architecture (BA, Mimar Sinan University, 1985), computer-aided architecture and design (MA, Straclyde University), and multimedia design (Mimar Sinan University, 1993). He has founded Yıldız Technical University’s Department of Communication Design, Interactive Media Design Graduate Program, and the PhD program in Art and Design. He continues to lead a team of interaction design researchers at the Koç University Design Lab.
\end{IEEEbiography}

\end{document}